\documentclass[12pt, aps,amsmath,showpacs,superscriptaddress]{revtex4-1}
\usepackage{graphicx,dcolumn,subfig,caption,bbm,amssymb,amsmath,ulem,indentfirst,amsthm,color}
\usepackage{natbib}
\begin{document}
\title{Comparing semiclassical mean-field and 1-exciton approximations in evaluating optical response under strong light-matter coupling conditions}

%\author{Bingyu Cui$^{1,2}$, Abraham Nitzan}
%\email{anitzan@sas.upenn.edu}
%\affiliation{Department of Chemistry, University of Pennsylvania, Philadelphia, Pennsylvania 19104,
%USA}
%\affiliation{School of Chemistry, Tel Aviv University, Tel Aviv 69978, Israel}
%\date{\today}

\author{Bingyu Cui}
\affiliation{Department of Chemistry, University of Pennsylvania, Philadelphia, Pennsylvania 19104,
USA}
\affiliation{School of Chemistry, Tel Aviv University, Tel Aviv 69978, Israel}

\author{Maxim Sukharev}
\affiliation{Department of Physics, Arizona State University, Tempe, Arizona 85287, USA}
\affiliation{College of Integrative Sciences and Arts, Arizona State University, Mesa, Arizona 85201, USA}

\author{Abraham Nitzan}
\email{anitzan@sas.upenn.edu}
\affiliation{Department of Chemistry, University of Pennsylvania, Philadelphia, Pennsylvania 19104,
USA}
\affiliation{School of Chemistry, Tel Aviv University, Tel Aviv 69978, Israel}

\date{\today}

\begin{abstract}
\noindent The rigorous quantum mechanical description of the collective interaction of many molecules with the radiation field is usually considered numerically intractable, and approximation schemes must be employed. Standard spectroscopy usually contains some levels of perturbation theory, but under strong coupling conditions, other approximations are used. A common approximation is the 1-exciton model in which processes involving weak excitations are described using a basis comprising of the ground state and singly excited states of the molecule cavity-mode system. In another frequently used approximation in numerical investigations, the electromagnetic field is described classically, and the quantum molecular subsystem is treated in the mean-field Hartree approximation with its wavefunction assumed to be a product of single molecules' wavefunctions. The former disregards states that take long time to populate, and is therefore essentially a short time approximation. The latter is not limited in this way, but by its nature, disregards some intermolecular and molecule-field correlations. In this work, we directly compare results obtained from these approximations when applied to several prototype problems involving the optical response of molecule-in-optical cavities systems. In particular, we show that our recent model investigation (J. Chem. Phys. 157, 114108 (2022)) of the interplay between the electronic strong coupling and molecular nuclear dynamics using the truncated 1-exciton approximation agrees very well with the semiclassical mean-field calculation.
\end{abstract}
\pacs{}
\maketitle

\section{Introduction}

The interaction of molecular systems with light in proximity to dielectric boundaries has long been a subject of study as a fundamental and applied science problem. Due to its importance, much effort is directed toward developing theoretical and numerical tools for such studies \cite{Feist2017,Cao2020,Climent2021,Li2022,Mandal2022,Sidler2022}. A full numerical treatment of interacting systems that comprise molecules, dielectric (usually metal) structures, and the quantum radiation field is usually beyond our reach, and standard approaches resort to various approximations \cite{Sukharev2017}: First, the electromagnetic field is often treated via classical Maxwell's equations. Second, the metal and other dispersive materials are described as continuum objects characterized by given dielectric response functions (models that use electron hydrodynamics coupled to Maxwell equations \cite{Scalora2010,Zeng2009,Sukharev2022b}, which make it possible to go beyond linear response are notable exceptions). Finally, the dynamics of molecular subsystems are described on the mean-field level, assuming that the total molecular wavefunction is the product of the wavefunctions of individual molecules. At the same time, the coupling between the molecules and the classical radiation field is handled on the Ehrenfest level, whereupon the radiation field responds to the expectation value of the local current induced in the molecular system. Further approximations involving the molecular subsystem, such as representing molecules as 2-electronic states systems coupled to a single nuclear coordinate, are also often made and even replacing the molecular medium with a dielectric continuum using, e.g., Lorentz oscillator model relating the dielectric function to molecular properties can provide useful insights. For a detailed description of this and related numerical procedures, see Ref. \cite{Sukharev2017}.

The validity of such models in situations involving many molecules interacting with metal plasmons or cavity modes excitations is not obvious. Collective molecular behavior is often implicated in the optical response of such systems, however, by the main approximations outlined above (a classical description of the radiation field, the mean-field (also referred to in the present context as time-dependent Hartree (TDH)) approximation for the molecular system and the Ehrenfest description of the interaction between them), which do not fully account for molecule-field and intermolecular correlations that may be important in such circumstances. An often-used alternative simplification that does not sacrifice such quantum correlations is the single exciton approximation (SEA), known in the quantum chemistry context as configuration interaction singles (CIS), in which the molecular quantum dynamics are described using a truncated basis that includes no more than one excited molecule. If molecular nuclear dynamics are included, additional truncation \cite{Cui2022}, or semiclassical nuclear dynamics \cite{Luk2017} has been used in the framework of the SEA, while in a recent paper \cite{sukharev2022} both classical radiation field and the TDH approximation for the molecular subsystem was used with a full quantum description of the molecular nuclear motion. We note in passing that in simulations of vibrational strong coupling, a fully classical description of the molecules and the cavity mode was found to be very useful \cite{Li2021,Li2022}.

This paper examines the applicability of these descriptions in simulations of molecules interacting with cavity modes. We focus on two models. First is the Tavis Cummings (TC) model \cite{Tavis1968,Tavis1969} for $N$ otherwise independent 2-level molecules interacting with a single cavity mode. The SEA for this model is often used to rationalize the main observation of polaritonic response in molecules positioned in optical cavities, i.e., the Rabi splitting that scales as $\sqrt{N}$ and corresponding Rabi oscillations in the time domain. Here we compare results based on the fully quantum SEA treatment of this model (referred to below as TC1) to calculations based on the classical treatment of the radiation field and the TDH approximation for the molecules (referred to below as TC2). The second model considered is an extension of the TC model recently used by us to examine the interplay between collective molecular response and molecular internal motions \cite{Cui2022}. The standard model that extends TC dynamics to include nuclear motions is the Holstein Tavis Cumming (HTC) model, in which the molecular subsystem includes $N$ molecules, each described by a 2-electronic state and one harmonic vibration with standard intramolecular vibronic interaction \cite{Holstein1959,Holstein1959b,Spano2015,Herrera2016,Galego2015,Galego2016b,Wu2016,Zeb2017}. In our model, the harmonic vibration is replaced by an intramolecular 2-state system with an interstate coupling that depends on the electronic state, making it possible to investigate the interplay between collective electronic dynamics in the cavity and the molecular internal 2-state dynamics. Here we compare our previously obtained results \cite{Cui2022} based on the fully quantum SEA with further truncation of the molecular internal basis (referred to below as the CN1 model) to a calculation based on the corresponding model with a classical cavity mode and the mean-field (TDH) description of the molecular dynamics that does not involve any basis truncation (CN2). Because the CN1 model was employed to examine how the molecular internal dynamics may influence the collective molecular electronic response \cite{Cui2022}, a test of the applicability of the mean-field description of this model is particularly important.

Details of these models are provided in Sec. II and in the Supplementary Information (SI), the results of our calculations are presented in Sec. III. Section IV summarizes our main observations and concludes.

\section{Models and methods}
In this section, we describe the models utilized to perform numerical comparisons between the 1-exciton and the classical radiation/molecular mean-field approximations: The first two are the Tavis-Cummings model restricted to the 1-exciton subspace (TC1) and its analog in which the cavity mode is described classically, the molecules are treated in the mean-field approximation and the Ehrenfest approximation is used to treat the field-molecule interaction (TC2). The others are the extended and similarly truncated TC model in which each molecule carries two inner states (CN1), and the corresponding model in which instead of basis truncation we employ a classical representation for the cavity mode, the mean-field approximation for molecules and the Ehrenfest approximation for their interaction (CN2).\\

The Tavis-Cummings (TC) Hamiltonian is \cite{Tavis1968,Tavis1969}
\begin{equation}
    \hat{H}_{TC}=\hbar\omega_c\hat{a}^\dagger\hat{a}+\hbar\sum_{j=1}^N\left[\omega_{xg}\hat{\sigma}^+_j\hat{\sigma}^-_j+\frac{g}{2}(\hat{a}^\dagger+\hat{a})(\hat{\sigma}^-_j+\hat{\sigma}^+_j)\right].
    \label{eq:HHTC}
\end{equation}
In Eq. \eqref{eq:HHTC}, the operator $\hat{a}$ ($\hat{a}^\dagger$) annihilates (creates) a cavity-mode photon of frequency $\omega_c$, while $\hat{\sigma}_j=|g_j\rangle\langle e_j|$ and $\hat{\sigma}^\dagger_j=|e_j\rangle\langle g_j|$ respectively affect the downward and upward transitions between the lower $|g_j\rangle$ and upper $|e_j\rangle$ electronic states of molecule $j$. $E_{xg}\equiv\hbar\omega_{xg}$, the molecular electronic transition energy, and the cavity-molecule coupling parameter $g$ are taken to be the same for all molecules. The disorder associated with the distribution of molecular orientations relative to the cavity modes is ignored. In the TC1 calculation the dynamics of this model is restricted to the 1-exciton subspace so that either the cavity mode is excited with all molecules in their ground electronic states, or only one molecule is excited while all others including the cavity are in their ground states \footnote{Note that in the 1-exciton approximation, we can replace the Hamiltonian \eqref{eq:HHTC} by its rotating state version,
\begin{equation}
    \hat{H}_{TC-RWA}=\hbar\omega_c\hat{a}^\dagger\hat{a}+\hbar\sum_{j=1}^N\left[\omega_{xg}\hat{\sigma}^+_j\hat{\sigma}^-_j+\frac{g}{2}(\hat{a}^\dagger\hat{\sigma}^-_j+\hat{a}\hat{\sigma}^+_j)\right],
    \label{eq:HHTCrwa}
\end{equation}
}. The system size (here and in the other models considered) is assumed to be much smaller than the radiation wavelength $2\pi c/\omega_c$ ($c$ is the speed of light). The basis of singly excited states is 
\begin{equation}
    |X_j\rangle=|e_j\rangle\prod_{k\neq j}|g_k\rangle,j=0,...,N,
    \label{eq:xj}
\end{equation}
where only the cavity mode $(j=0)$ or a single molecule ($j=1,...,N$) is excited. In terms of the subset of molecular singly excited states, the molecular bright state is 
\begin{equation}
    |B\rangle=\frac{1}{\sqrt{N}}\sum_{j=1}^N|X_j\rangle.
    \label{eq:B}
\end{equation}
The interaction of the molecular bright state with the cavity singly excited state leads to the Rabi splitting between the lower and upper polariton states \cite{Li2022},
\begin{equation}
    \Omega_R\sim g\sqrt{N},
    \label{eq:rabi}
\end{equation}
whose observation is characterized as "strong coupling".

Consider now the analogous model in the semiclassical mean-field approximation (TC2). The cavity mode is represented by a classical harmonic oscillator so that operators $\hat{a}$ and $ \hat{a}^\dagger$ become their classical analogs - superpositions of position $x$ and momentum $p$,
\begin{align}
        a&=\frac{1}{2}\left(\sqrt{\frac{2\omega_c}{\hbar}}x-i\sqrt{\frac{2}{\hbar\omega_c}p}\right),\\
        a^*&=\frac{1}{2}\left(\sqrt{\frac{2\omega_c}{\hbar}}x+i\sqrt{\frac{2}{\hbar\omega_c}p}\right).
\end{align}
The corresponding Hamiltonian reads
\begin{align}
    \label{eq:semiTC}
    H_{TC2}&=\frac{p^2+\omega_c^2x^2}{2}+\hbar\sum_{j=1}^N\left[\omega_{xg}\hat{\sigma}^+_j\hat{\sigma}^-_j+\frac{g}{2}\sqrt{\frac{2\omega_c}{\hbar}}x(\hat{\sigma}_j^++\hat{\sigma}_j^-)\right].
\end{align}
We make the time-dependent Hartree approximation by writing the molecular wavefunction $\Psi$ as a product of wavefunctions $\psi$ of individual molecules, i.e.,
\begin{subequations}
\begin{align}
    \Psi(t)&=\prod_{j=1}^N\psi_j(t),\\
    |\psi_j(t)\rangle&=c_{jg}(t)|g_j\rangle+c_{je}(t)|e_j\rangle.
    \label{eq:waveftrunc}
\end{align}
\end{subequations}
The coefficients $c$ evolve in time according to (see Sec. I in the SI for the detailed derivation)
%\begin{equation}
%    i\hbar\dot{\psi}_j=\left(\frac{p^2+\omega_cx^2}{2N}\right)\psi_j+\hat{H}_j\psi_j+\hat{V}_j\psi_j,
%    \label{eq:singleH}
%\end{equation}
%with 
%\begin{align}
%    \hat{H}_j&\equiv\omega_{xg}\hat{\sigma}^+_j\hat{\sigma}^-_j,\\
%    \hat{V}_j&\equiv\frac{g}{2}\sqrt{\frac{2\omega_c}{\hbar}}x(\hat{\sigma}_j^++\hat{\sigma}_j^-).
%\end{align}
\begin{align}
    i\hbar\frac{d}{dt}\left(\begin{matrix}
    c_{jg}\\
    c_{je}
   \end{matrix}\right)=
    \left(\begin{matrix}
    E_g+\frac{p^2+\omega_c^2x^2}{2N} &\frac{g}{2}\sqrt{\frac{2\omega_c}{\hbar}}x \\
\frac{g}{2}\sqrt{\frac{2\omega_c}{\hbar}}x &E_e+\frac{p^2+\omega_c^2x^2}{2N}
\end{matrix}\right)\left(\begin{matrix}
    c_{jg}\\
    c_{je}
    \end{matrix}\right).
    \label{eq:singleH}
\end{align}
Clearly, the total wavefunction $\Psi$ is normalized as long as the normalization condition holds for each $\psi_j$. The evolution of the classical oscillator is obtained by applying the Ehrenfest theorem, leading to
\begin{align}
   \begin{cases}
   \dot{p}=-\omega_c^2x-\frac{g}{2}\sqrt{\frac{2\omega_c}{\hbar}}\sum_{j=1}^N\left[\langle\psi_j|\hat{\sigma}_j^-|\psi_j\rangle+\langle\psi_j|\hat{\sigma}_j^+|\psi_j\rangle\right],\\
   \dot{x}=p,
   \end{cases}
   \label{eq:classicN}
\end{align}
where 
\begin{align}
   \langle\psi_j|\hat{\sigma}_j^-|\psi_j\rangle=\langle\psi_j|\hat{\sigma}_j^+|\psi_j\rangle^*=c_{jg}^*c_{je}.
   \label{eq:classicalcoe}
\end{align}
Using Eqs. \eqref{eq:singleH} and \eqref{eq:classicN}, it can be shown that the time evolution according to this mixed quantum-classical dynamics conserves the total energy (see Sec. I in the SI for details).

One possible extension of the TC model to account for the nuclear motion is the Holstein-Tavis-Cummings (HTC) model \cite{Holstein1959,Holstein1959b} in which each two-electronic-level molecule is associated with a single (nuclear) harmonic mode with vibronic coupling represented by the standard polaron model. In Ref. \cite{Cui2022}, we have proposed a simplified version of this model in which the harmonic modes are replaced with two-state entities with interstate coupling $\lambda$ that depends on the electronic state (and is therefore the analogue of the vibronic coupling of the HTC model). Denoting the inner states of molecule $j$ by $|a_j\rangle$ and $|b_j\rangle$, and their energy spacing $\hbar\Delta\omega$, the Hamiltonian is
\begin{equation}
    \hat{H}_{CN}=\hbar\omega_c\hat{a}^\dagger\hat{a}+\hbar\sum_{j=1}^N\left[\omega_{xg}\hat{\sigma}^+_j\hat{\sigma}^-_j+\frac{g}{2}(\hat{\sigma}^-_j\hat{a}^\dagger+\hat{\sigma}^+_j\hat{a})+\Delta\omega\hat{\tau}_j^+\hat{\tau}_j^-+\lambda\hat{\sigma}_j^+\hat{\sigma}_j^-(\hat{\tau}_j^++\hat{\tau}_j^-)\right].
      \label{eq:ETC}
\end{equation}
where $\hat{\tau}_j^+=|b_j\rangle\langle a_j|,\hat{\tau}_j^-=|a_j\rangle\langle b_j|$. In Ref. \cite{Cui2022} we have used this model to investigate the interplay between the vibronic (internal) coupling and collective optical response of the molecular system in optical cavities. 

In the fully quantum truncated basis approximation to the dynamics of this Hamiltonian (the CN1 model) \cite{Cui2022}, we use the following consideration: The complete Hilbert space of this Hamiltonian is spanned by states $|XV\rangle$ where $X$ and $V$ correspond to the electronic and internal ("nuclear") molecular subspaces. In the spirit of the Born approximation, we assume that the molecule-radiation field interaction does not couple different internal states. It follows \cite{Cui2022} that if the initial molecular internal state is $|V_0\rangle=\prod_{j=1}^N|a_j\rangle$, namely, all molecules start in internal state $a$ (analog of "all molecules start in the ground state of their vibrational mode" in the HTC model), then for a time long relative to the inverse of the Rabi splitting Eq. \eqref{eq:rabi} and short relative to $g^{-1}$, it is sufficient to consider a truncated internal basis that includes only states $|V_0\rangle$ and $|V_j\rangle=|b_j\rangle\prod_{k\neq j}|a_k\rangle$, disregarding states in which more than one molecule has changed its internal state. A general state in this truncated basis will be denoted $|j,k\rangle=|X_j\rangle|V_k\rangle,j,k=0,...,N$.

Next, consider the same model \eqref{eq:ETC} in the semiclassical mean-field approximation (the CN2 model). The cavity mode is now represented as a classical oscillator, so the Hamiltonian \eqref{eq:ETC} becomes
\begin{align}
    \hat{H}_{CN2}=\frac{p^2+\omega_c^2x^2}{2}+\hbar\sum_{j=1}^N\left[\omega_{xg}\hat{\sigma}^+_j\hat{\sigma}^-_j+\frac{g}{2}\sqrt{\frac{2\omega_c}{\hbar}}x(\hat{\sigma}_j^++\hat{\sigma}_j^-)+\Delta\omega\hat{\tau}_j^+\hat{\tau}_j^-+\lambda\hat{\sigma}_j^+\hat{\sigma}_j^-(\hat{\tau}_j^++\hat{\tau}_j^-)\right].
    \label{eq:semiETC}
\end{align}
In addition, the truncated basis approximation is now replaced by the mean-field TDH approximation: 
\begin{subequations}
\begin{align}
|\Psi(t)\rangle&=\prod_{j=1}|\psi_j(t)\rangle,\\
|\psi_j\rangle&=c_{jga}(t)|g_j,a_j\rangle+c_{jgb}(t)|g_j,b_j\rangle+c_{jea}(t)|e_j,a_j\rangle+c_{jeb}(t)|e_j,b_j\rangle.
\label{eq:wavefsemi}
\end{align}
\end{subequations}
The time evolution of coefficients $c$ is given by
\begin{align}
    i\hbar\frac{d}{dt}\left(\begin{matrix}
    c_{jga}\\
    c_{jgb}\\
    c_{jea}\\
    c_{jeb}
    \end{matrix}\right)=
    \left(\begin{matrix}
    E_g+\frac{p^2+\omega_c^2x^2}{2N} &0 &\frac{g}{2}\sqrt{\frac{2\omega_c}{\hbar}}x
&0 \\
0 &E_g+\frac{p^2+\omega_c^2x^2}{2N}+\hbar\Delta\omega &0 &\frac{g}{2}\sqrt{\frac{2\omega_c}{\hbar}}x\\
\frac{g}{2}\sqrt{\frac{2\omega_c}{\hbar}}x &0 &E_e+\frac{p^2+\omega_c^2x^2}{2N} &\lambda\\
0 &\frac{g}{2}\sqrt{\frac{2\omega_c}{\hbar}}x &\lambda &E_e+\frac{p^2+\omega_c^2x^2}{2N}+\hbar\Delta\omega
\end{matrix}\right)\left(\begin{matrix}
    c_{jga}\\
    c_{jgb}\\
    c_{jea}\\
    c_{jeb}
    \end{matrix}\right),
\end{align}
while equations of motion for the classical cavity oscillator are obtained by using the Ehrenfest theorem as before. This leads again to Eq. \eqref{eq:classicN}, with Eq. \eqref{eq:classicalcoe} now replaced by
\begin{align}
   \langle\psi_j|\hat{\sigma}_j^-|\psi_j\rangle=\langle\psi_j|\hat{\sigma}_j^+|\psi_j\rangle^*=c_{jga}^*c_{jea}+c_{jgb}^*c_{jeb}.
   \label{eq:classicalcoeinternal}
\end{align}

The performance of these models is examined under three dynamical scenarios. First, starting from the molecular bright state Eq. \eqref{eq:B} in the TC1 and CN1 models or $|B\rangle|V_0\rangle$ \footnote{In the mean-field representation used in models TC2 and CN2, the molecular bright state is taken by letting $c_{je}=N^{-1/2}, c_{jg}=(1-N^{-1})^{1/2}$ in Eq. \eqref{eq:waveftrunc} or $c_{jea}=N^{-1/2},c_{jga}=(1-N^{-1})^{1/2}$ in Eq. \eqref{eq:wavefsemi}.} (referred to as the $a$-bright state) for the TC2 and CN2 models, the system is evolved according to $|\Psi(t)\rangle=e^{-i\hat{H}t/\hbar}|\Psi(0)\rangle=\sum_{j=0}^N\sum_{k=0}^Nc_{jk}(t)|j,k\rangle$ by diagonalizing the Hamiltonian in the truncated basis models (TC1 and CN1) or by solving the semiclassical TDH equations (\eqref{eq:semiTC} for the TC2 model and \eqref{eq:semiETC} for the CN2 model). Similarly, we denote it the $a$-ground state when molecules are in the ground electronic and inner $a$ state ($|X_0\rangle|V_0\rangle$ in the CN1 model and $\prod_j|g_j,a_j\rangle$ in the CN2 model). The TC1 and TC2 models are compared by evaluating the Rabi splitting and the dynamics of the energy exchange between the molecules and the cavity mode. In comparing the CN1 and CN2 models we focus on the internal dynamics. We denote by $d_0(t)$ the probability that the molecular system excited at time $t$ remains in its initial inner state $|V_0\rangle$ in which all molecules are in their internal state $a$. Similarly, $d_1(t)$ is the probability that at time $t$ the molecular system is in the state where the excited molecule has changed its inner state from $a$ to $b$. In the CN1 model, these probabilities are given by
\begin{align}
    d_0(t)&=\sum_{j=1}^N|c_{j0}(t)|^2,\notag\\
    d_1(t)&=\sum_{j=1}^N|c_{jj}(t)|^2,
    \label{eq:dt}
\end{align}
whereas in the CN2 model, we have
\begin{align}
    d_0(t)&=\sum_j|c_{jea}(t)|^2\prod_{k\neq j}|c_{kga}(t)|^2,\notag\\
    d_1(t)&=\sum_j|c_{jeb}(t)|^2\prod_{k\neq j}|c_{kga}(t)|^2.
    \label{eq:semidt}
\end{align}
Note that, when the system is propagated from the $a$-bright state, most molecular states remain unexcited and in inner state $a$, so the product $\prod_{k\neq j}|c_{kga}|^2$ is very close to unity and can be ignored.

In the second scenario, the transient dynamics is initiated by a short Gaussian pulse, 
\begin{subequations}
\begin{equation}
F(t)=\exp(-(t/\tau_0)^2),
\label{eq:Ft}
\end{equation}
incident on the ground state system. In this ground state, each molecule is assumed to be in the $a$ internal state and that this state is not changed by the optical interaction. In the CN1 model, this is done by adding a driving level (a level whose population remains constant) that is coupled to the excited cavity state by coupling 
\begin{equation}
V=WF(t)(|G,0\rangle\langle0,0|+|0,0\rangle\langle G,0|),
\label{eq:V}
\end{equation}
in the cavity and 
\begin{equation}
V=WF(t)|G,0\rangle
\label{eq:V2}
\end{equation}
\end{subequations}
outside it. Here we denote by $|G,0\rangle$ the states where the cavity and all molecules are in the ground state with all molecules in the inner state $a$ and, as before, by $|B,0\rangle$ the molecular $a$-bright state. In the CN2 model, the equation of motion for the classical oscillator, Eq. \eqref{eq:classicN}, is augmented by a term $A\cos(\omega t)F(t)$ where $\omega=\omega_{xg}$ for exciting the $a$-bright state outside the cavity and $\omega=\omega_{xg}-\Omega_R/2$ (in order to excite the lower polariton). Note that, the parameters $A$ and $W$ are related: The additional term in the Hamiltonian Eq. \eqref{eq:semiETC} is $Ax$, implying that $W=A\langle 0|x|1\rangle=A\sqrt{\hbar/(2\omega_c)}$ (where $0$ and $1$ are states of the cavity mode).

Finally, in the third scenario, the steady state behavior of these models is considered by adding driving and damping processes to the systems described by the Hamiltonian \eqref{eq:HHTC}, \eqref{eq:semiTC}, \eqref{eq:ETC}, \eqref{eq:semiETC}. Specifically, driving is applied to the cavity mode. The details of these modifications are again somewhat different in the quantum TC1 and CN1 models and in their semiclassical TC2 and CN2 counterparts, and are described next below. In the TC1 and CN1 models this is done as described in Ref. \cite{Cui2022} by adding a driving level of energy $\hbar\omega$ (that corresponds to a photon of frequency $\omega$ "seated" on the ground state) that is coupled to the $a$-cavity mode $|X_0\rangle|V_0\rangle$ (inside the cavity) or to the molecular $a$-bright state, $|B\rangle|V_0\rangle$ (outside the cavity), with a coupling $W$ that is taken small enough so that the response to this driving is linear. Additionally, each state in the excited state manifold of the uncoupled system (the eigenstates of Hamiltonian of Eq. \eqref{eq:ETC} in the absence of the $g$ and $\lambda$ terms) is assumed to be coupled to its own broad band continuum, implying that the evolution equations for these states are modified by adding damping terms. For example, for the CN1 model, we can add damping rates $-\eta_a/2$ to the evolution of the amplitude of the $|X_j\rangle|V_0\rangle$ state and $-\eta_b/2$ to the evolution of $|X_j\rangle|V_k\rangle,j=1,...,N$ states so that if a state $|s\rangle$ is damped ($s=a,b$), the equation for the corresponding coefficient $c_s$ is supplemented according to $dc_s/dt=...-i\eta_sc_s/2$ and is associated with the corresponding damping flux $d|c_s|^2/dt=...-\eta_s|c_s|^2$ (see Sec. II in SI for details). We often take $\eta_a=0$ to describe the physics in which the transition $a\rightarrow b$ in a molecule is followed by an irreversible process of rate $\eta_b$ to form some product \footnote{Taking $\eta_a\neq 0$ corresponds to relaxation processes that do not lead to this product}. In addition, a damping parameter $\eta_c$ is imposed in the same way on all states in which the cavity mode is excited. It represents all possible decay channels of this mode: reflection, transmission and dissipation in the cavity boundaries. In terms of the Green’s "function" associated with the Hamiltonian of Eq. \eqref{eq:ETC} that has been modified in the following way,
\begin{equation}
    \mathbf{G}=\frac{1}{\omega\mathbf{I}-\mathbf{H}+i\pmb{\eta}/2},
    \label{eq:G}
\end{equation}
where $\mathbf{I}$ is the unit matrix and $\pmb{\eta}$ is the diagonal matrix of damping coefficients $\eta_s$, this leads \cite{Cui2022} to the following expression for the absorption lineshape
\begin{equation}
    L(\omega)\sim-\text{Im}G_{00,00}(\omega).
    \label{eq:quantumlineshape}
\end{equation}
Also, assuming that the decay via the $b$ states leads to an observable product, the yield of product formation in the CN1 model is given by \cite{Cui2022} 
\begin{align}
    Y_b(\omega)&=-\eta_b\frac{\sum_{j=1}^N|G_{00,jj}|^2}{\text{Im}G_{00,00}}.
    \label{eq:quantumyield}
\end{align}
The functions $L(\omega)$ and $Y_b(\omega)$ are the observables addressed by this calculation.

To describe similar observables in the TC2 and CN2 models we add an oscillatory driving force of frequency $\omega$ and amplitude $A$ and a dissipation term $\eta_c$ to the classical oscillator equations of motion so that Eq. \eqref{eq:classicN} becomes 
\begin{align}
   \dot{p}+\omega_c^2x+\frac{g}{2}\sqrt{\frac{2\omega_c}{\hbar}}\sum_{j=1}^N\left[\langle\psi_j|\hat{\sigma}_j^-|\psi_j\rangle+\langle\psi_j|\hat{\sigma}_j^+|\psi_j\rangle\right]=f(t)-\eta_c p; \quad f(t)=A\cos(\omega t),
   \label{eq:psourcedamp}
\end{align}
where $\eta_c\langle p^2\rangle$ (average over a period $2\pi/\omega$) is the energy dissipation flux, $\langle W\rangle_T$, out of the classical cavity mode. We also add damping terms to the molecular dynamics as described above. In particular for the CN2 model, i.e. Eq. \eqref{eq:semiETC}, we impose $\eta_a$ and $\eta_b$ relaxation rates on the states $|e_j,a_j\rangle$ and $|e_j,b_j\rangle$ respectively (we have found that $\eta_a\neq0$ needs to be taken to avoid the occurrence of singularities but it can be taken small enough to be physically irrelevant). The semiclassical equations of motion are evolved until the steady state is reached. The absorption lineshape is now calculated from the mechanical work (averaged over a period) done at this steady state on the classical cavity oscillator 
\begin{equation}
    L\sim\langle p(t)f(t)\rangle=(2\pi/\omega)^{-1}\int_{t_0}^{t_0+2\pi/\omega} p(t)f(t)dt,
    \label{eq:semilineshape}
\end{equation}
while, as defined before, the product yield is given by
\begin{align}
    \label{eq:semiyield}
    Y_b=\frac{E_{xg}\langle\sum_j\eta_b|c_{jeb}|^2\rangle}{\langle p(t)f(t)\rangle}.
\end{align}
Note that unlike Eq. \eqref{eq:quantumlineshape}, Eq. \eqref{eq:semilineshape} represents the energy absorption rate as a function of frequency, which is why $E_{xg}$ appears in the numerator of Eq. \eqref{eq:semiyield}. For the same reason, the results from Eq. \eqref{eq:quantumlineshape} and Eq. \eqref{eq:semilineshape} can be compared only by their shapes and not by their absolute numerical values.

It should be noted that these steady state flux calculations stand in apparent contrast to the underlying assumption of short time validity of the truncated basis approximation used in models TC1 and CN1. In Ref. \cite{Cui2022} we have argued that the truncated basis may still work well in such situations provided that the damping is fast enough so that during the lifetime of the excited molecule/cavity system no appreciable population can be accumulated in states that are not included in the truncated basis. The semiclassical mean-field models are not similarly restricted, so an agreement between results from these two very different approximations will provide important support to their validity. Note that the quantities defined in Eqs. \eqref{eq:quantumlineshape}-\eqref{eq:semiyield} should be evaluated from the long time (steady state) solution of the dynamical equations of motion so that the initial conditions for such a calculation are unimportant.

In addition to the three scenarios described above for which we compared the performance of the truncated 1-exciton and the semiclassical mean-field models, our semiclassical approach has been also used to observe the transient dynamic during and following the switching of the excitation by the exciting pulse. To this end, we replace the constant amplitude $A$ by the time-dependent function $AS(t)$ where 
\begin{align}
    S(t)=
    \begin{cases}
    0, t<0,\\
    \sin^2(\pi t/t_s),0\leq t\leq t_s/2,\\
    1, t>t_s/2,
    \end{cases}
    \label{eq:envelope}
\end{align}
where $t_s$ is the switching time, so that $f(t)$ in Eq. \eqref{eq:psourcedamp} is replaced by $f(t)=AS(t)\cos(\omega t)$. Following the start of the process at $t=0$ we define the following observables 
\begin{subequations}
\begin{align}
    L_a(t)&=E_{xg}\eta_a\sum_{j=1}^N\int_0^{t}|c_{jea}(t')|^2dt',\\
    L_b(t)&=E_{xg}\eta_b\sum_{j=1}^N\int_0^{t}|c_{jeb}(t')|^2dt',\\
    L_c(t)&=\eta_c\int_0^{t}p(t')^2dt',\\
    Y_b(t)&=\frac{L_b}{L_a+L_b+L_c},
    \label{eq:fluxyield}
\end{align}
\label{eq:switching}
\end{subequations}
which are used to gain insights into the transient response of the system.

\section{Results}
In the calculation presented in this section, we focus only on the zero detuning case, taking the cavity mode frequency equal to the molecular electronic transition frequency, $\omega_c=\omega_{xg}$ and set $\hbar=1$. In what follows, time and frequency are shown in units of $g^{-1}$ and $g$ respectively.

\begin{figure}[!htp]
\subfloat[][]{
\begin{minipage}[t]{0.5\textwidth}
\flushleft
\includegraphics[width=0.8\textwidth]{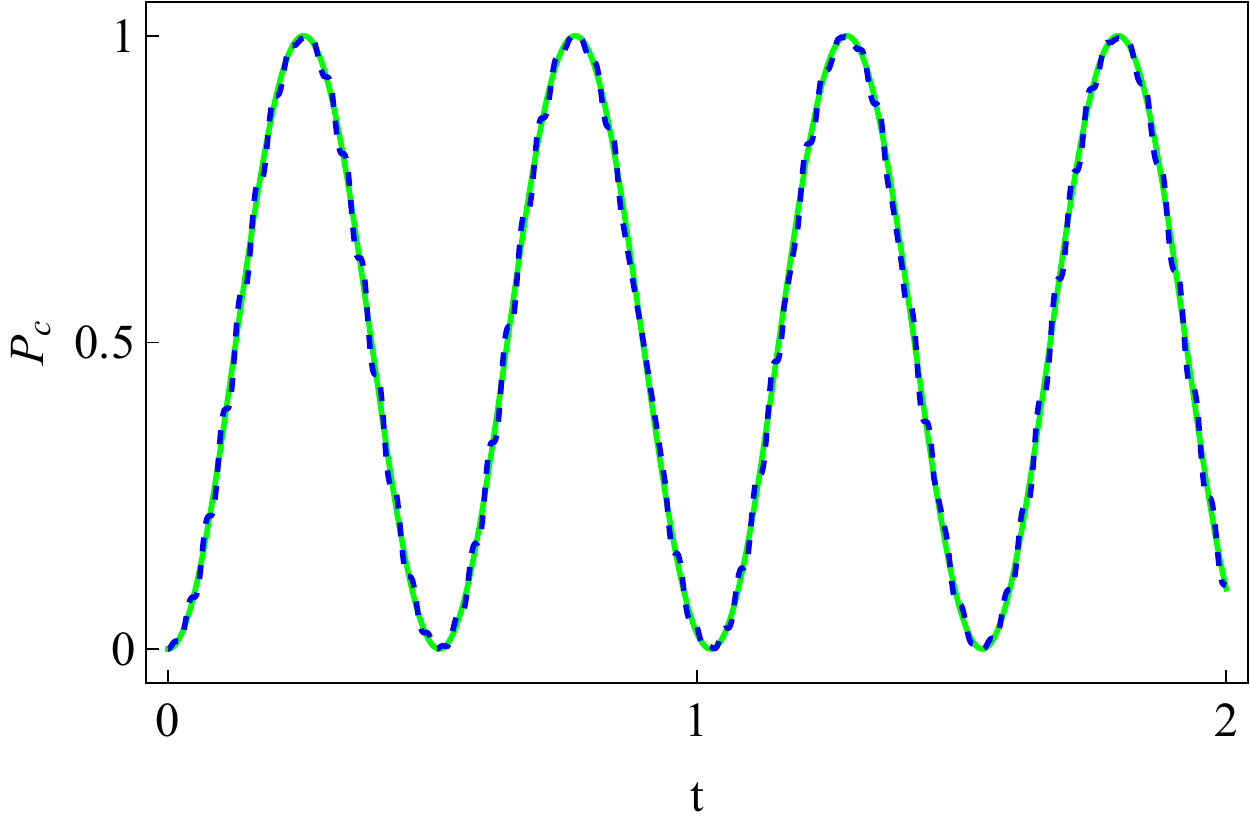}
\end{minipage}
}
\\
\subfloat[][]{
\begin{minipage}[t]{0.5\textwidth}
\flushleft
\includegraphics[width=0.8\textwidth]{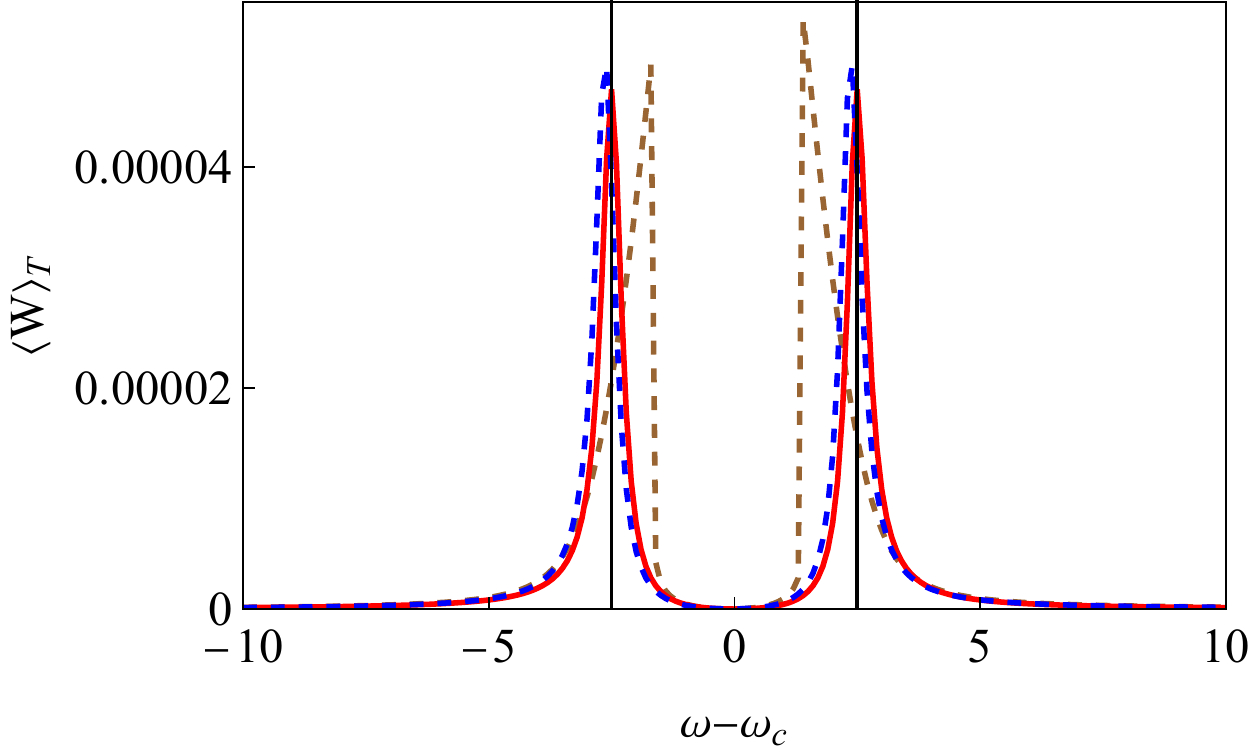}
\end{minipage}
}\\
\subfloat[][]{
\begin{minipage}[t]{0.5\textwidth}
\flushleft
\includegraphics[width=0.8\textwidth]{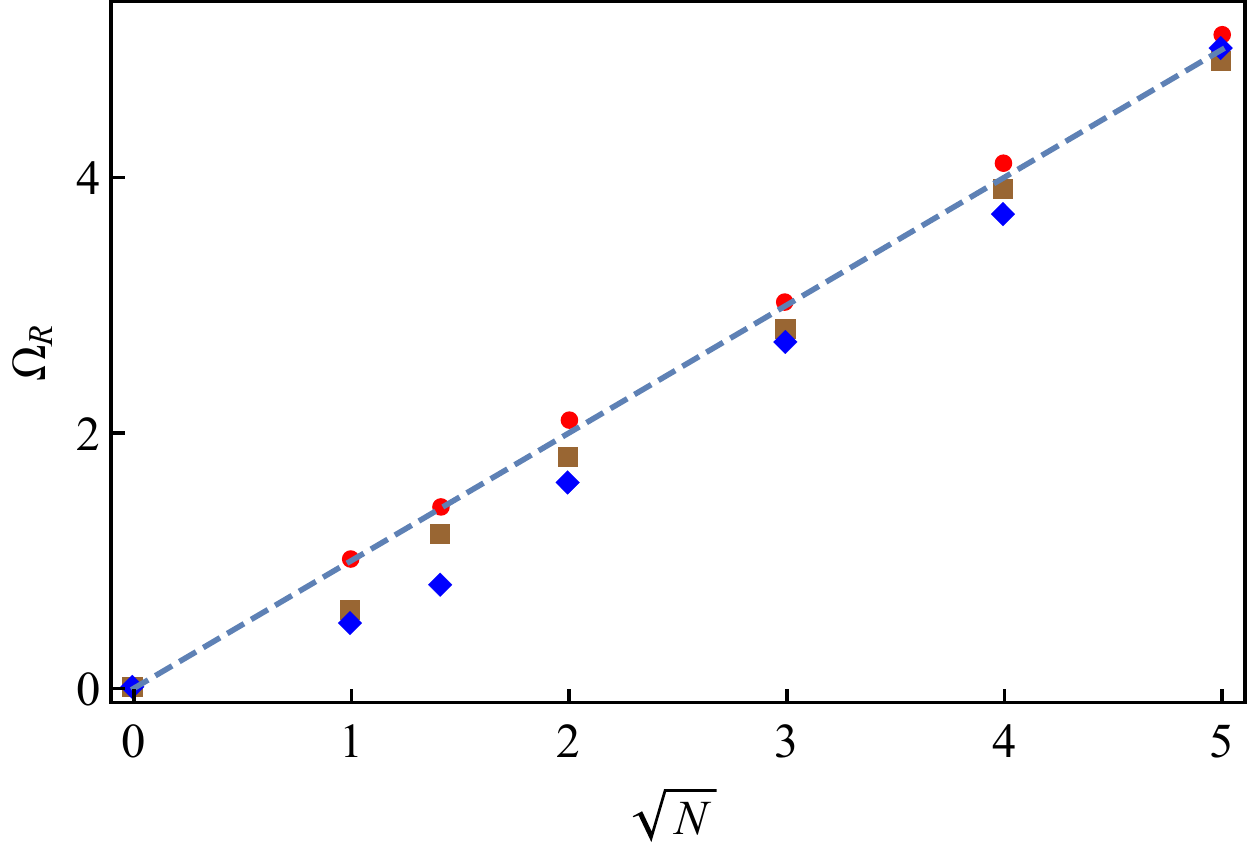}
\end{minipage}
}
\caption{Dynamics and lineshapes calculated for the TC1 and TC2 models. Panel (a): the time evolution of the cavity mode population calculated by the TC1 (green solid line) or TC2 (blue dashed line) models with the initial state taken to be the molecular bright state $|B\rangle$. The size of the molecular cluster is $N=150$. Other parameters are $\hbar\omega_c=E_{xg}=100g$. The initial state of the cavity mode is the quantum ground state in the TC1 model and the zero energy state, $p(0)=0,x(0)=0$, in the TC2 model. Panel (b): the average power (arbitrary units) dissipation in the TC1 (blue dashed lines) and TC2 (red solid lines) models for a system consisting of $N=25$ atoms coupled to the cavity mode. Vertical lines label the polariton frequencies calculated for the TC1 model: $\omega_c\pm g\sqrt{N}/2$. Parameters are $g=0.1, \hbar\omega_c=E_{xg}=100g,\eta_c=g, W=0.01g, A=0.01$. The brown dashed line is the result of the TC2 model with a larger amplitude $A=2$. The left peaks of the three curves (red, blue, brown) are scaled to the same height. Panel (c): the Rabi splitting, calculated as the distance between the peaks obtained in the lineshape calculations such as seen in panel (b), plotted against the square root of the number of molecules, with various pumping amplitudes $A=$0.01 (red dots), 0.1 (brown dots) and 0.15 (blue dots). The linear dashed line is the analytical result for the TC1 model, with a gradient $g$. All energy/time values are displayed in units of $g/g^{-1} (\hbar=1)$.}
\label{fig:1}
\end{figure}

Figure \ref{fig:1} shows results obtained using the TC1 and TC2 models. Panel (\ref{fig:1}a) shows the time evolution of the population of the cavity mode ($\langle\Psi(t)|\hat{a}^\dagger\hat{a}|\Psi(t)\rangle$ in TC1 (full green line), $|a(t)|^2$ in TC2 (dashed blue line)) starting from the bright mode of the molecular system, $|\Psi(t=0)\rangle=|B\rangle$. The agreement between the resulting Rabi oscillations (the analytical result of the TC1 model is $P_{c}(t)=\sin^2(\Omega_R t/2)$) indicates the validity of both approximations for the chosen system parameters (see Fig. \ref{fig:1} caption). Panel (b) shows the steady state absorption spectrum calculated for the two models under different pumping conditions. For the TC2 model with the parameters used, the steady state is reached at about $t_0=30/\eta_c$ ($\eta_c$, introduced in Eq. \eqref{eq:psourcedamp}, is the damping coefficient associated with the cavity mode). In the TC1 model, the linear response calculation (see Ref. \cite{Cui2022}) is not sensitive to the pumping amplitude $A$. For the TC2 mode, the direct numerical calculation is not restricted to linear response, and the dependence on $A$ shows the effect of deviation from this limit. It should be emphasized however that the validity of the semiclassical mean-field approximation may also depend on the pumping strength. Panel (c) shows the Rabi splitting displayed against $\sqrt{N}$  obtained from the steady state absorption lineshape. Again, for a small pumping amplitude, the results of two approximations are nearly identical, showing the familiar linear dependence on $\sqrt{N}$, while deviations from this linearity are seen for larger pumping amplitude $A$ in the TC2 model.

\begin{figure}[!htp]
\includegraphics[width=0.8\textwidth]{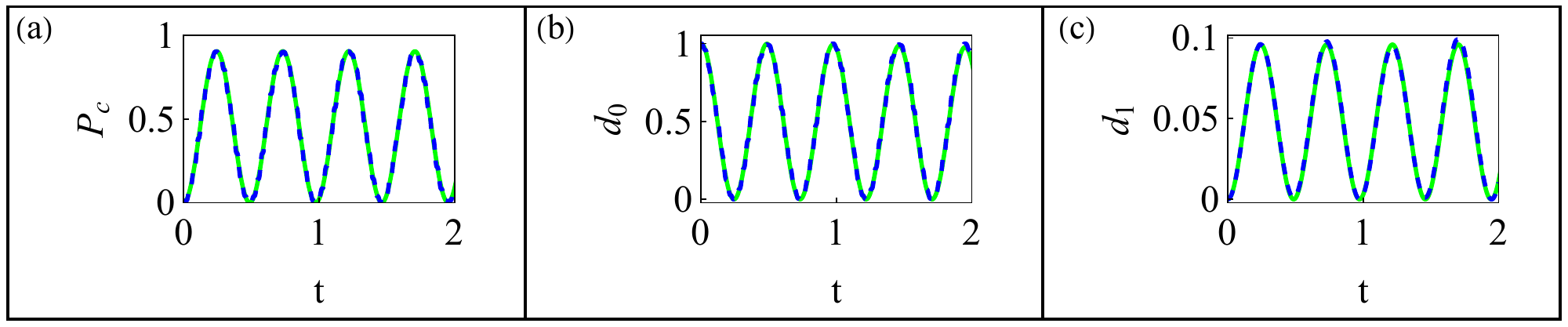}
\caption{Time evolution in the CN1 (green solid lines) and CN2 (blue dashed lines) models. The molecular cluster size is $N=150$ and each molecule has two inner states characterized by energy separation $\Delta\omega=0$ and interstate coupling $\lambda=2g$. Other parameters are $g=0.1, \hbar\omega_c=\hbar\omega_{xg}=100g$. The initial state is the $a$-bright state $\Psi(t=0)=|B\rangle|V_0\rangle$. Panel (a) shows the energy population in the cavity mode, while panels (b) and (c) show respectively the quantities $d_0$ and $d_1$ defined in Eq. \eqref{eq:dt} for the quantum model CN1 and in Eq. \eqref{eq:semidt} for the semiclassical mean-field model CN2. }
\label{fig:2}
\end{figure}

\begin{figure}[!htp]
\includegraphics[width=0.8\textwidth]{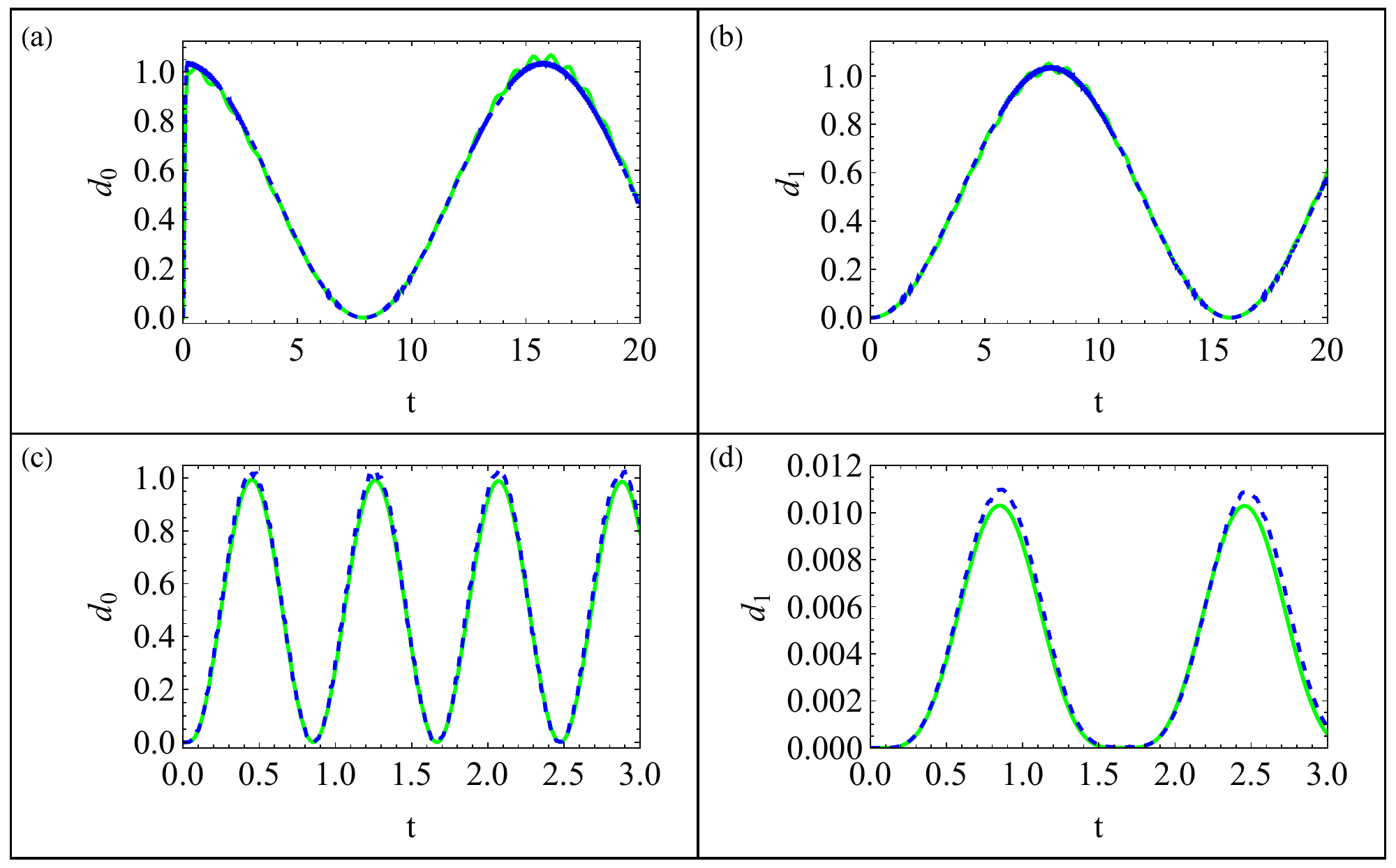}
\caption{Time evolution in the CN1 (green solid lines) and CN2 (blue dashed lines) models following the short pulse. Before the pulse (see text for details), $N=60$ molecules are in the $a$-ground state. Panels (a-b) show time evolution outside the cavity; panels (c-d) represent situations when molecules are inside the cavity. The quantities $d_0$ and $d_1$ are defined in Eq. \eqref{eq:dt} for the quantum model CN1 or in Eq. \eqref{eq:semidt} for the semiclassical mean-field model CN2. The frequency of the external field is $\omega_{xg}$ (outside the cavity) or $\omega_{xg}-\Omega_R/2$ (inside the cavity) and the characteristic time of the pulse envelope $\tau_0$ is taken $\pi/(5\Omega_R)$. Other parameters are $g=0.1, \hbar\omega_c=E_{xg}=100g, A=0.01, \Delta\omega=0$ and $\lambda=0.2g$. The heights of all curves are scaled such that $d_0$ oscillates between $0$ and $1$.}
\label{fig:3}
\end{figure}

Figure \ref{fig:2} shows some characteristics of the dynamical evolution based on models CN1 and CN2, starting from the $a$-bright state $|B\rangle|V_0\rangle$ of the molecular subsystem. Shown is the evolution of the cavity mode population (calculated as Figure 1), displaying the characteristic Rabi oscillations (panel (a)) as well as the evolution of the cumulative populations $d_0$ and $d_1$ (Eqs. \eqref{eq:dt} and \eqref{eq:semidt}) pertaining to the molecular internal dynamics. Nearly perfect agreement between the quantum 1-exciton model and the semiclassical mean-field approximation is again observed on the displayed timescale.

Similar results for models CN1 and CN2 are shown in Fig. \ref{fig:3}, for a system that starts in the $a$-ground state following a pulse excitation, Eqs. \eqref{eq:Ft}-\eqref{eq:V2}, with the pulse time $\tau_0=\pi/(5\Omega_R)$. Panels (a) and (b) show results for a molecular system outside the cavity while panels (c) and (d) show the corresponding results inside the cavity. Excellent agreement is again seen between the results obtained for the two models.

\begin{figure}[!htp]
\includegraphics[width=0.8\textwidth]{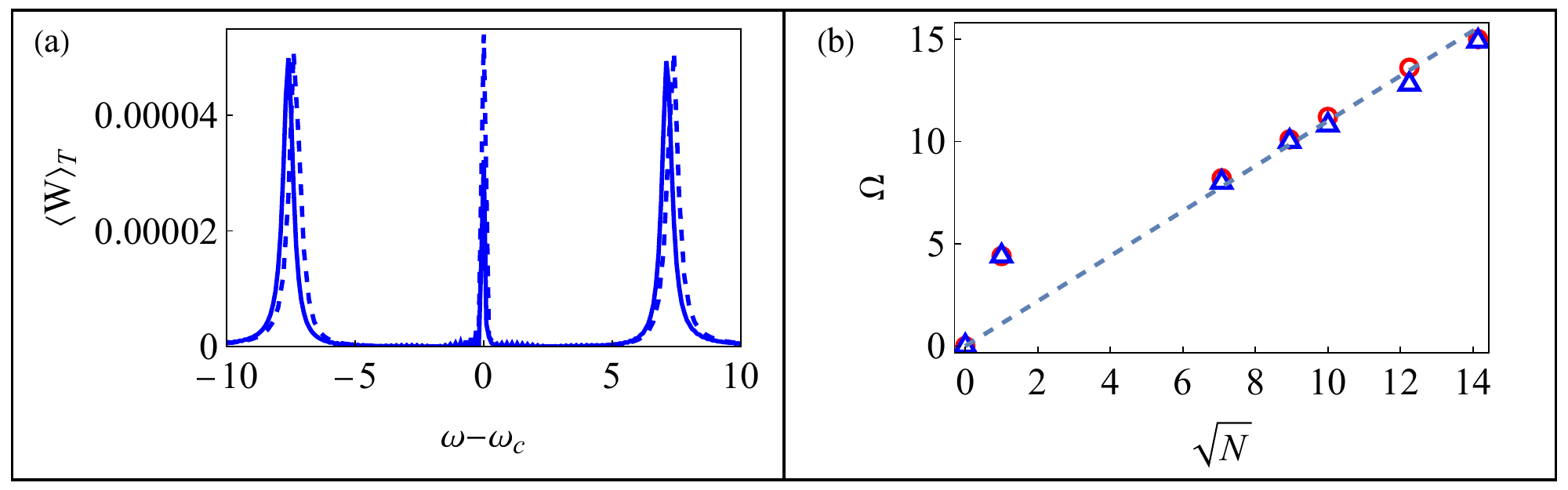}
\caption{Lineshape (arbitrary units) and the separation between the two outer peaks calculated for models CN1 and CN2. Panel (a): the average power dissipation out of the cavity mode plotted against the pumping frequency obtained from the CN1 (solid line) and CN2 (dashed line) models of a system consisting of $N=200$ atoms, using the parameters $g=0.1, \hbar\omega_c=E_{xg}=100g,\eta_c=g,\eta_a=\eta_b=0, W=0.01g, A=0.01, \Delta\omega=0$ and $\lambda=2g$. The left peaks of the two curves (dashed and solid) are scaled to the same height. For the CN2 model using these parameters, the steady state is reached at about $t_0=30/\eta_c$. Panel (b): the separation between two outer peaks obtained from model CN1 (red circles) and CN2 (blue triangles) plotted against the square root of the number of molecules. The linear fit has a gradient of $1.1g$.}
\label{fig:4}
\end{figure}

\begin{figure}[!htp]
\subfloat[][]{
\begin{minipage}[t]{0.5\textwidth}
\flushleft
\includegraphics[width=0.8\textwidth]{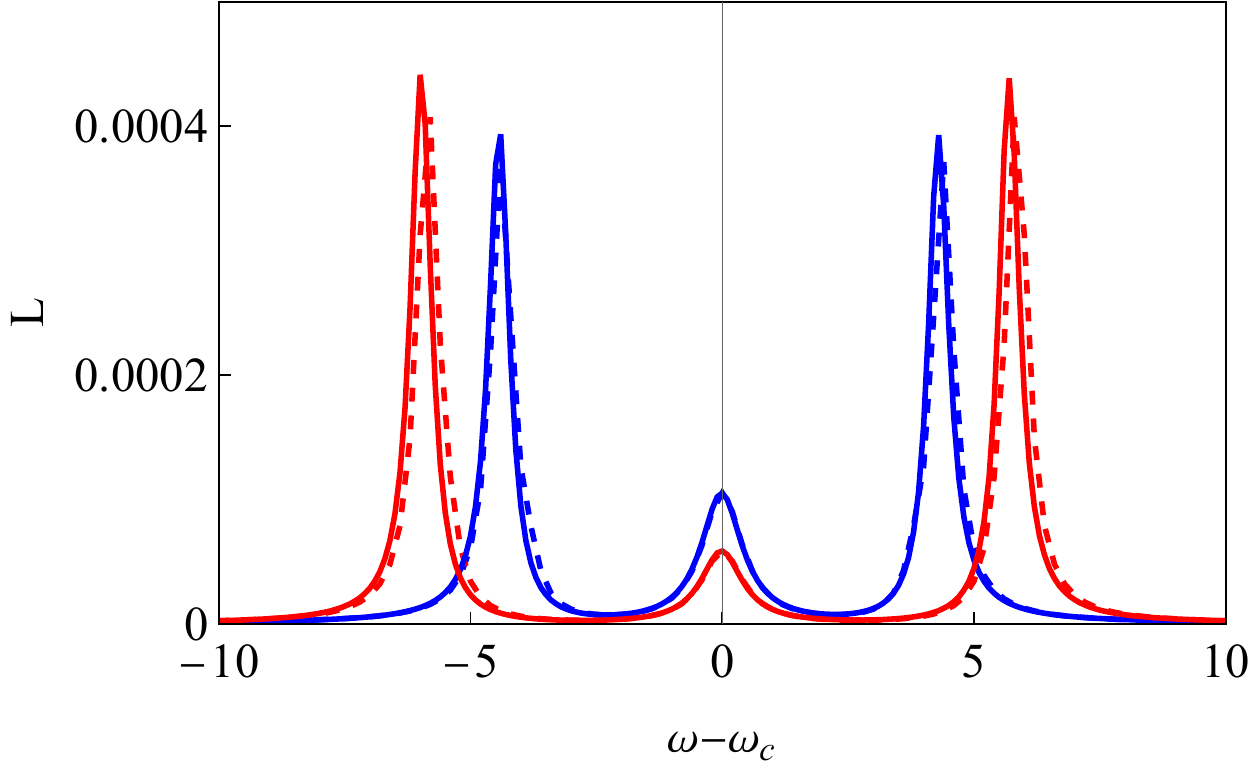}
\end{minipage}
}
\\
\subfloat[][]{
\begin{minipage}[t]{0.5\textwidth}
\flushleft
\includegraphics[width=0.8\textwidth]{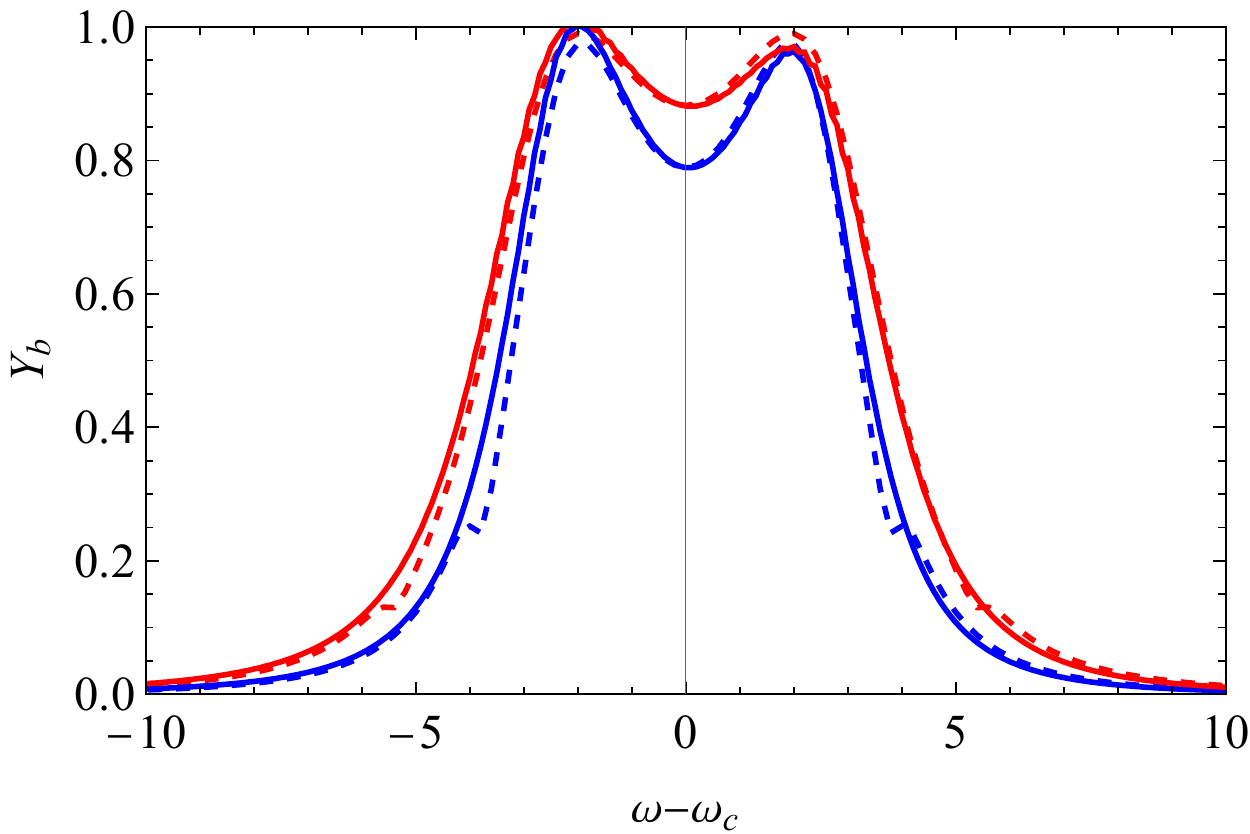}
\end{minipage}
}
\caption{(a) Absorption lineshapes (arbitrary units), plotted against the pumping frequency. The lineshapes calculated from Eqs. \eqref{eq:quantumlineshape} and \eqref{eq:semilineshape} are scaled to have the same height of the middle peak. (b) The yield $Y_b$, calculated from the CN1 (Eq. (\ref{eq:quantumyield}), dashed lines), and CN2 (Eq. (\ref{eq:semiyield}), solid lines) models, for molecular clusters of size $N=60$ (blue) or $N=120$ (red). Damping channels are opened for excited molecules in inner state $b$ and for the cavity mode/classical oscillator with damping coefficients $\eta_b=\eta_c=g$, respectively, while $\eta_a=0$. Other parameters are $g=0.1,\hbar\omega_c=E_{xg}=100g,W=0.01g, A=0.01, \Delta\omega=0$ and $\lambda=2g$.}
\label{fig:5}
\end{figure}

\begin{figure}[!htp]
\subfloat[][]{
\begin{minipage}[t]{0.5\textwidth}
\flushleft
\includegraphics[width=0.8\textwidth]{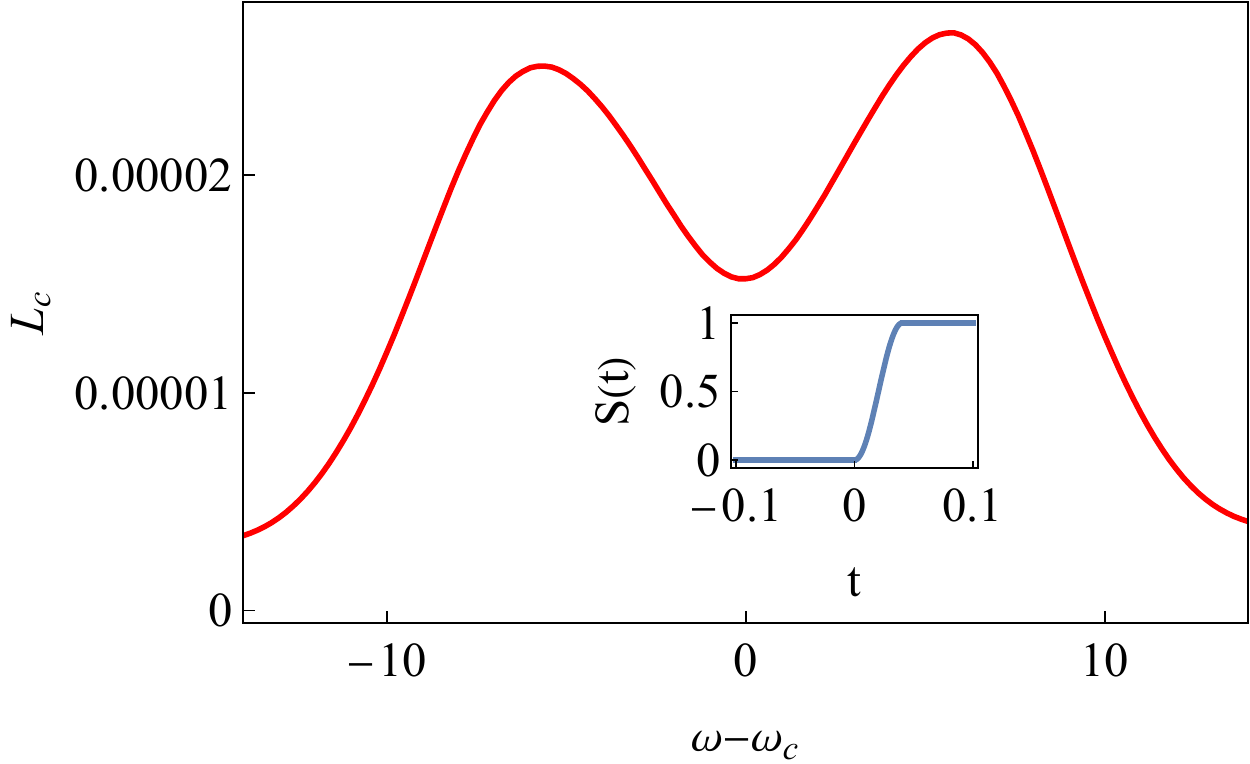}
\end{minipage}
}
\\
\subfloat[][]{
\begin{minipage}[t]{0.5\textwidth}
\flushleft
\includegraphics[width=0.8\textwidth]{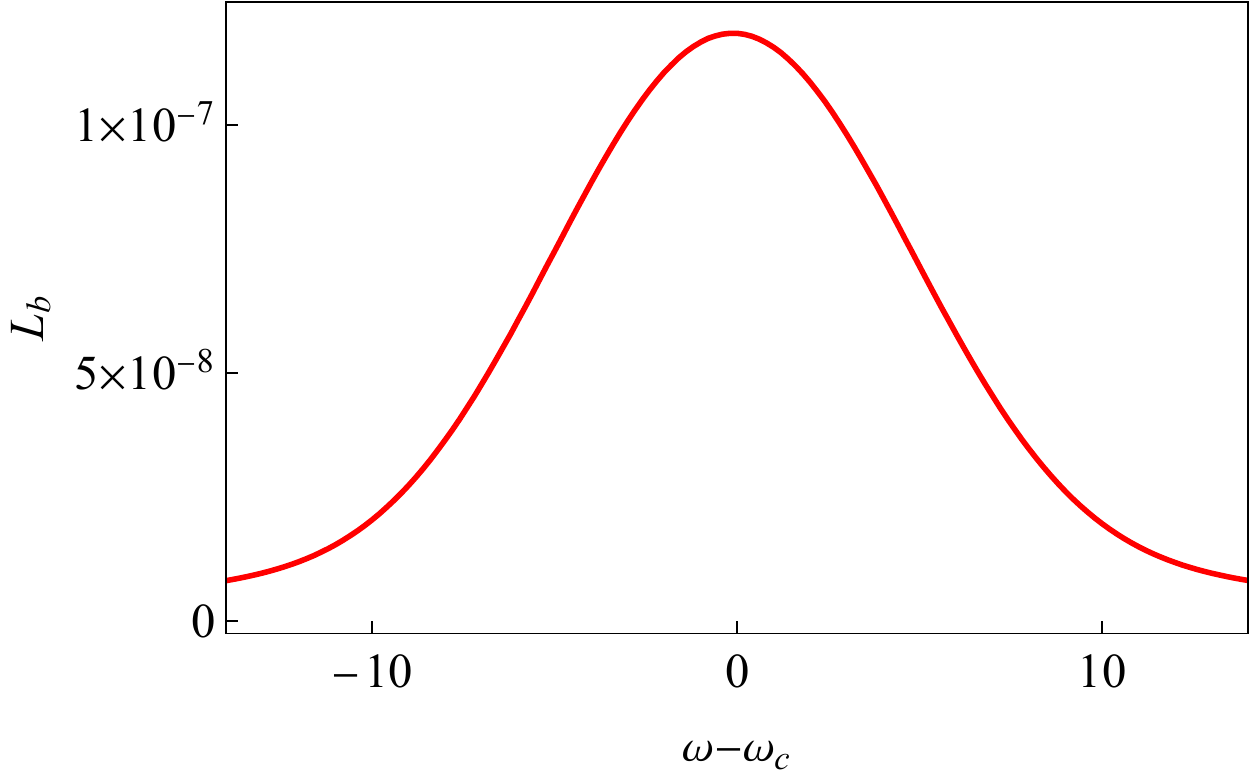}
\end{minipage}
}
\\
\subfloat[][]{
\begin{minipage}[t]{0.5\textwidth}
\flushleft
\includegraphics[width=0.8\textwidth]{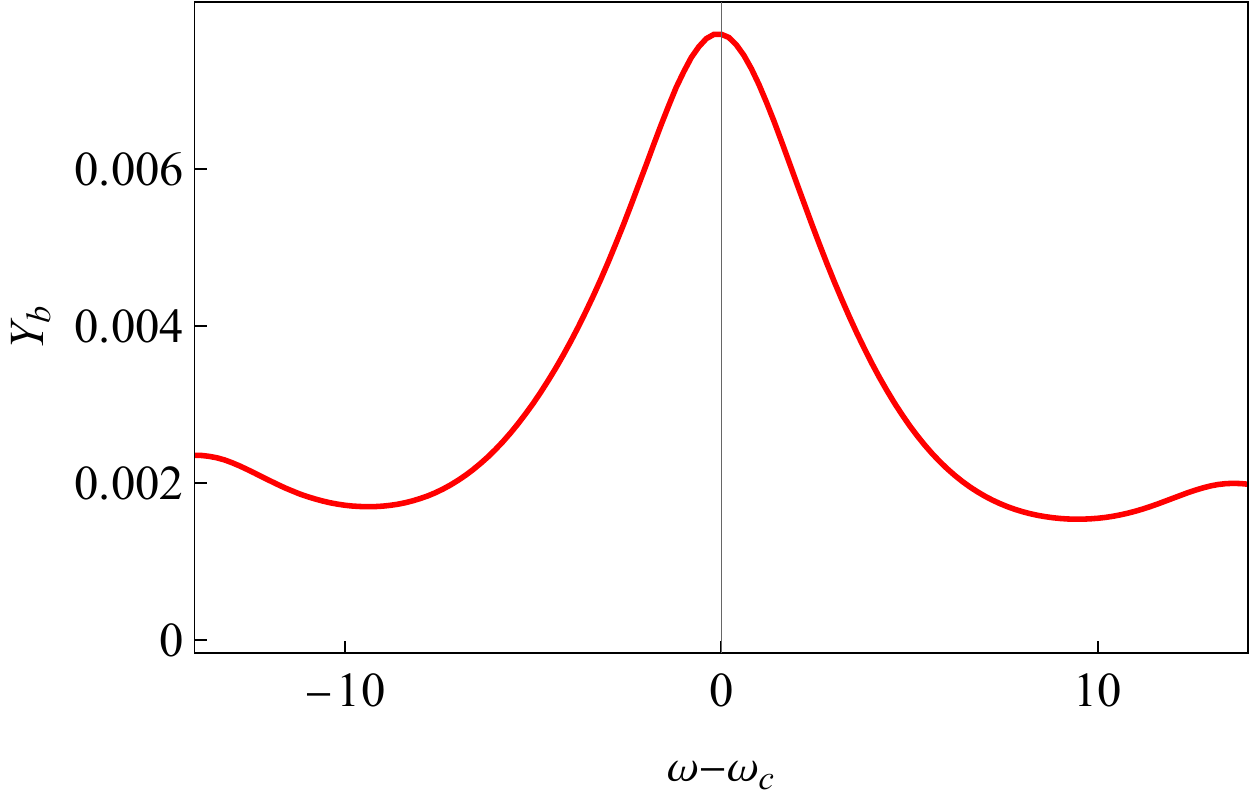}
\end{minipage}
}
\caption{Fluxes and yield obtained for the CN2 model when the system is under the switching field. Panel (a): flux through the cavity mode represented by the classical oscillator, the inset is the shape of the transient field, c.f., Eq. \eqref{eq:envelope}; panel (b): flux through inner states $b$ in excited molecules; panel (c): yield of inner states $b$, c.f., Eq. \eqref{eq:fluxyield}. Before the switching, there are $N=60$ molecules in the $a$-ground state. The characteristic time of the pulse envelope $t_s$ is $\pi/(5g\sqrt{N})$. The damping coefficients are $\eta_c=\eta_b=0.5g, \eta_a=0$. The snapshot is taken at $t=10t_s$. Other parameters are $g=0.1,\hbar\omega_c=E_{xg}=100g, A=0.01, \Delta\omega=0$ and $\lambda=0.2g$.}
\label{fig:6}
\end{figure}

Steady state lineshape calculated for models CN1 and CN2 are compared in Figs. \ref{fig:4} and \ref{fig:5}, which show steady state fluxes out of specified channels, as well as flux ratios for yield calculations (see details below) for continuous pumping of molecules that start in the $a$-ground state. We emphasize again that this test is more stringent because the truncated-basis CN1 model is essentially a short time that may remain valid at a long time only under certain conditions specified as discussed in Section II. The lineshapes obtained for the molecules-in-cavity system are characterized by three peaks: with only the flux through the cavity mode we would see the standard Rabi splitting with two peaks, which here corresponds to the outer two peaks in Figs. \ref{fig:4}(a) and \ref{fig:5}(a). The additional center peak is due to the b-states that do not shift because populating a b-state creates a molecular state that is not strongly coupled to the cavity. In Fig. \ref{fig:4} steady state is achieved by constantly pumping the ground state system while imposing dissipation on the cavity mode $(\eta_a=\eta_b=0, \eta_c\neq 0)$. The steady state lineshape and the separation $\Omega$ between the two outer peaks, calculated for models CN1 and CN2 are compared in Fig. \ref{fig:4}(a) and Fig. \ref{fig:4}(b), respectively. $\Omega$, essentially the splitting between the $a$-bright state and the cavity mode, is identical to the Rabi splitting $\Omega_R$ in the limit $\lambda=0$. For finite $\lambda$, the linear dependence of the Rabi splitting on the square root of the size of the molecular cluster is recovered only for large $N$ when $g\sqrt{N}>\lambda$.

Similar steady state results are shown in Fig. \ref{fig:5} for a different loss scheme: $\eta_a=0,\eta_b=\eta_c=g$ (more results for $\eta_a=\eta_b=g,\eta_c=0$ are shown in the SI  (Fig. S2)). Again, these calculations generally show good agreement between the quantum truncated basis calculations (model CN1) and the semiclassical mean-field results (model CN2). A small qualitative disagreement is worth noticing: The very small blips in the quantum yield observed at polaritonic frequencies in the CN1 model are not reproduced by the CN2 calculation. In Ref. \cite{Cui2022} we have speculated that these blips may be related to observations made in Ref. \cite{sukharev2022} of slower rates of molecular dissociation at the polariton frequencies, but we now need to conclude that these blips are probably artefacts indicating the limitation of the truncated basis application to the steady state situation as discussed above and in Ref. \cite{Cui2022}. Indeed the slowing down of product formation observed in Ref. \cite{sukharev2022} is a transient effect that we now reproduce in Fig. \ref{fig:6}.

A study of molecular dissociation following electronic excitation of molecules inside a Fabry–P\'{e}rot cavity under strong coupling conditions indicates that following the onset of optical pumping there is a transient slowdown of molecular dissociation at the polaritonic resonance frequencies \cite{sukharev2022}. Figure \ref{fig:6} shows the result of our attempt to reproduce this observation using the CN2 model. To this end, we use the scheme described by Eqs. (\ref{eq:envelope}-\ref{eq:switching}): the incident field is switched on during the time $t_s$ and the total flux out of $b$-states is calculated at a later time $t$ in the presence of a competing process-damping of the cavity mode (a similar calculation for the case where the competing flux is out of states $a$ is shown as Fig. S2 in SI). The population flux out of states $b$ and the energy flux out of the cavity mode are shown as functions of the incident frequency $\omega$ in Figs. \ref{fig:6}(a-b) and the corresponding yield, Eq. \eqref{eq:fluxyield}, is shown in Fig. \ref{fig:6}(c). Interestingly, dips, albeit shallow, are seen near the polariton frequencies. In Ref. \cite{sukharev2022} we have interpreted these dips as manifestations of the local distortion of the dissociation potential surface due to the collective polariton formation. In analogy, in the present model when the $a$-polariton is excited the energy separation between $a$ and $b$ state is larger than at other pumping frequencies because of the polariton Rabi shift, leading to a slower $a$ to $b$ transition rate.

\section{Discussion and conclusion}
In this paper, we have compared the performance of two common approximations used in the studies of molecular ensembles coupled to optical fields. One, often used in theoretical modeling, is the single exciton approximation, where in models such as the Tavis-Cummings (TC) \cite{Tavis1968,Tavis1969} and Holstein-Tavis-Cummings (HTC) \cite{Holstein1959,Holstein1959b} the dynamics of the subspace associated with a single photon absorption is investigated (often with additional basis truncation that disregards states that are not populated during times of interest). The other, frequently used in numerical simulations, is based on a semiclassical approach in which the electromagnetic field is evolved using classical Maxwell's equations while the Schr\"{o}dinger or Liouville equations are used to describe the molecular dynamics on the mean-field (Hartree) level. This mixed quantum-classical dynamics makes it necessary to use also the Ehrenfest approximation for the coupling between the quantum and classical dynamics. These very different approximations (one is essentially a short time approximation and the other disregards possibly important correlations), were compared using the TC model as well as a simplified HTC model in which the molecular oscillator is replaced by a 2-level system characterized by similar internal timescale and coupling. Excellent agreements between the two approximation schemes were obtained for low excitation levels, indicating the validity of both approximations in this regime. Another important observation is that the semiclassical mean-field approximation accounts well for the manifestations of collective behavior explored in this study: the collective Rabi splitting and oscillations as well as the implications of collective polaritonic energy shift on reaction dynamics.

The success of the mean-field approximation in accounting for the collective response associated with the bright state of the molecular ensemble can be rationalized by the following observation: In this approximation, the quantum state of the molecular subsystem for 2-state molecules is $|\Psi\rangle=\prod_j(c_{jg}|g_j\rangle+c_{je}|e_j\rangle)=\prod_jc_{jg}|g_j\rangle+\sum_kc_{ke}|e_k\rangle\prod_{j\neq k}c_{jg}|g_j\rangle+...$ where $|g_j\rangle$ and $|e_j\rangle$ are the ground and excited states of molecule $j$ and where higher order terms may be disregarded if $c_{je}\ll c_{jg}$ for all $j$. Thus, the single exciton subspace and its collective characteristics are fully described in this limit of the mean-field calculation. Furthermore, a single mode classical oscillator coupled to the bright mode of this quantum system can describe the essential characteristics of the Rabi splitting phenomenon as a model comprising two coupled classical oscillators shows \cite{Trm2014}. It is perhaps more surprising that the mean-field semiclassical pictures can account well for manifestations of collective behavior in the molecular internal (nuclear) dynamics. We attribute this success to the following observations. In the simple 2-internal state model used here and in Ref. \cite{Cui2022}, the Hartree molecular wavefunction is $|\Psi\rangle=\prod_j(c_{jga}|g_j,a_j\rangle+c_{jea}|e_j,a_j\rangle+c_{jgb}|g_j,b_j\rangle+c_{jeb}|e_j,b_j\rangle)$. This level of approximation can therefore account for an essential feature: A molecular internal $a\leftrightarrow b$ dynamics is fully correlated with the electronic state of \textit{that molecule}. Starting from the $a$-bright state in which all molecules are in internal states $a$, such $a\leftrightarrow b$ dynamics transforms it to a dark state which in a cavity environment can have significant energetic consequences as discussed in Ref. \cite{Cui2022} and \cite{sukharev2022}.

The success of the semiclassical mean-field approximation, as demonstrated here at least for low excitation levels, brings up the question of the limitations of this level of description. In particular, the fact that important aspects of collective response are successfully described on this level of approximation naturally leads us to question the possible limits of this success. This question bears on the recent interest in applications of such collective behaviors to the operation of devices such as envisioned "quantum batteries" \cite{Campaioli2018,Bhattacharjee2021}. We will address these questions in a future publication.

\section*{Supplementary Material}
See supplementary material for discussions on conditions to apply the mean-field approximation, details on adding damping terms in the TC1 and CN1 models and more figures complimentary for results presented in Sec. III.

\section*{Acknowledgements}
This material is based upon work supported by the U.S. National Science Foundation under Grant CHE1953701. M.S. is grateful for the support from the Air Force Office of Scientific Research under grant No. FA9550-22-1-0175.

\section*{Author Declarations}
\subsection*{Conflict of interest}
The authors have no conflicts to disclose. 

\section*{Data availability}
The data that support the findings of this study are available within the article [and its supplementary material].

\bibliography{reference}

\begin{thebibliography}{30}%
\makeatletter
\providecommand \@ifxundefined [1]{%
 \@ifx{#1\undefined}
}%
\providecommand \@ifnum [1]{%
 \ifnum #1\expandafter \@firstoftwo
 \else \expandafter \@secondoftwo
 \fi
}%
\providecommand \@ifx [1]{%
 \ifx #1\expandafter \@firstoftwo
 \else \expandafter \@secondoftwo
 \fi
}%
\providecommand \natexlab [1]{#1}%
\providecommand \enquote  [1]{``#1''}%
\providecommand \bibnamefont  [1]{#1}%
\providecommand \bibfnamefont [1]{#1}%
\providecommand \citenamefont [1]{#1}%
\providecommand \href@noop [0]{\@secondoftwo}%
\providecommand \href [0]{\begingroup \@sanitize@url \@href}%
\providecommand \@href[1]{\@@startlink{#1}\@@href}%
\providecommand \@@href[1]{\endgroup#1\@@endlink}%
\providecommand \@sanitize@url [0]{\catcode `\\12\catcode `\$12\catcode
  `\&12\catcode `\#12\catcode `\^12\catcode `\_12\catcode `\%12\relax}%
\providecommand \@@startlink[1]{}%
\providecommand \@@endlink[0]{}%
\providecommand \url  [0]{\begingroup\@sanitize@url \@url }%
\providecommand \@url [1]{\endgroup\@href {#1}{\urlprefix }}%
\providecommand \urlprefix  [0]{URL }%
\providecommand \Eprint [0]{\href }%
\providecommand \doibase [0]{http://dx.doi.org/}%
\providecommand \selectlanguage [0]{\@gobble}%
\providecommand \bibinfo  [0]{\@secondoftwo}%
\providecommand \bibfield  [0]{\@secondoftwo}%
\providecommand \translation [1]{[#1]}%
\providecommand \BibitemOpen [0]{}%
\providecommand \bibitemStop [0]{}%
\providecommand \bibitemNoStop [0]{.\EOS\space}%
\providecommand \EOS [0]{\spacefactor3000\relax}%
\providecommand \BibitemShut  [1]{\csname bibitem#1\endcsname}%
\let\auto@bib@innerbib\@empty
%</preamble>
\bibitem [{\citenamefont {Feist}\ \emph {et~al.}(2017)\citenamefont {Feist},
  \citenamefont {Galego},\ and\ \citenamefont {Garcia-Vidal}}]{Feist2017}%
  \BibitemOpen
  \bibfield  {author} {\bibinfo {author} {\bibfnamefont {J.}~\bibnamefont
  {Feist}}, \bibinfo {author} {\bibfnamefont {J.}~\bibnamefont {Galego}}, \
  and\ \bibinfo {author} {\bibfnamefont {F.~J.}\ \bibnamefont {Garcia-Vidal}},\
  }\href {\doibase 10.1021/acsphotonics.7b00680} {\bibfield  {journal}
  {\bibinfo  {journal} {{ACS} Photonics}\ }\textbf {\bibinfo {volume} {5}},\
  \bibinfo {pages} {205} (\bibinfo {year} {2017})}\BibitemShut {NoStop}%
\bibitem [{\citenamefont {Cao}\ \emph {et~al.}(2020)\citenamefont {Cao},
  \citenamefont {Cogdell}, \citenamefont {Coker}, \citenamefont {Duan},
  \citenamefont {Hauer}, \citenamefont {Kleinekath\"{o}fer}, \citenamefont
  {Jansen}, \citenamefont {Man{\v{c}}al}, \citenamefont {Miller}, \citenamefont
  {Ogilvie}, \citenamefont {Prokhorenko}, \citenamefont {Renger}, \citenamefont
  {Tan}, \citenamefont {Tempelaar}, \citenamefont {Thorwart}, \citenamefont
  {Thyrhaug}, \citenamefont {Westenhoff},\ and\ \citenamefont
  {Zigmantas}}]{Cao2020}%
  \BibitemOpen
  \bibfield  {author} {\bibinfo {author} {\bibfnamefont {J.}~\bibnamefont
  {Cao}}, \bibinfo {author} {\bibfnamefont {R.~J.}\ \bibnamefont {Cogdell}},
  \bibinfo {author} {\bibfnamefont {D.~F.}\ \bibnamefont {Coker}}, \bibinfo
  {author} {\bibfnamefont {H.-G.}\ \bibnamefont {Duan}}, \bibinfo {author}
  {\bibfnamefont {J.}~\bibnamefont {Hauer}}, \bibinfo {author} {\bibfnamefont
  {U.}~\bibnamefont {Kleinekath\"{o}fer}}, \bibinfo {author} {\bibfnamefont
  {T.~L.~C.}\ \bibnamefont {Jansen}}, \bibinfo {author} {\bibfnamefont
  {T.}~\bibnamefont {Man{\v{c}}al}}, \bibinfo {author} {\bibfnamefont
  {R.~J.~D.}\ \bibnamefont {Miller}}, \bibinfo {author} {\bibfnamefont {J.~P.}\
  \bibnamefont {Ogilvie}}, \bibinfo {author} {\bibfnamefont {V.~I.}\
  \bibnamefont {Prokhorenko}}, \bibinfo {author} {\bibfnamefont
  {T.}~\bibnamefont {Renger}}, \bibinfo {author} {\bibfnamefont {H.-S.}\
  \bibnamefont {Tan}}, \bibinfo {author} {\bibfnamefont {R.}~\bibnamefont
  {Tempelaar}}, \bibinfo {author} {\bibfnamefont {M.}~\bibnamefont {Thorwart}},
  \bibinfo {author} {\bibfnamefont {E.}~\bibnamefont {Thyrhaug}}, \bibinfo
  {author} {\bibfnamefont {S.}~\bibnamefont {Westenhoff}}, \ and\ \bibinfo
  {author} {\bibfnamefont {D.}~\bibnamefont {Zigmantas}},\ }\href {\doibase
  10.1126/sciadv.aaz4888} {\bibfield  {journal} {\bibinfo  {journal} {Science
  Advances}\ }\textbf {\bibinfo {volume} {6}} (\bibinfo {year} {2020}),\
  10.1126/sciadv.aaz4888}\BibitemShut {NoStop}%
\bibitem [{\citenamefont {Climent}\ \emph {et~al.}(2021)\citenamefont
  {Climent}, \citenamefont {Garcia-Vidal},\ and\ \citenamefont
  {Feist}}]{Climent2021}%
  \BibitemOpen
  \bibfield  {author} {\bibinfo {author} {\bibfnamefont {C.}~\bibnamefont
  {Climent}}, \bibinfo {author} {\bibfnamefont {F.~J.}\ \bibnamefont
  {Garcia-Vidal}}, \ and\ \bibinfo {author} {\bibfnamefont {J.}~\bibnamefont
  {Feist}},\ }in\ \href {\doibase 10.1039/9781839163043-00343} {\emph {\bibinfo
  {booktitle} {Theoretical and Computational Chemistry Series}}}\ (\bibinfo
  {publisher} {Royal Society of Chemistry},\ \bibinfo {year} {2021})\ pp.\
  \bibinfo {pages} {343--393}\BibitemShut {NoStop}%
\bibitem [{\citenamefont {Li}\ \emph {et~al.}(2022)\citenamefont {Li},
  \citenamefont {Cui}, \citenamefont {Subotnik},\ and\ \citenamefont
  {Nitzan}}]{Li2022}%
  \BibitemOpen
  \bibfield  {author} {\bibinfo {author} {\bibfnamefont {T.~E.}\ \bibnamefont
  {Li}}, \bibinfo {author} {\bibfnamefont {B.}~\bibnamefont {Cui}}, \bibinfo
  {author} {\bibfnamefont {J.~E.}\ \bibnamefont {Subotnik}}, \ and\ \bibinfo
  {author} {\bibfnamefont {A.}~\bibnamefont {Nitzan}},\ }\href {\doibase
  10.1146/annurev-physchem-090519-042621} {\bibfield  {journal} {\bibinfo
  {journal} {Annual Review of Physical Chemistry}\ }\textbf {\bibinfo {volume}
  {73}},\ \bibinfo {pages} {43} (\bibinfo {year} {2022})}\BibitemShut {NoStop}%
\bibitem [{\citenamefont {Mandal}\ \emph {et~al.}(2022)\citenamefont {Mandal},
  \citenamefont {Taylor}, \citenamefont {Weight}, \citenamefont {Koessler},
  \citenamefont {Li},\ and\ \citenamefont {Huo}}]{Mandal2022}%
  \BibitemOpen
  \bibfield  {author} {\bibinfo {author} {\bibfnamefont {A.}~\bibnamefont
  {Mandal}}, \bibinfo {author} {\bibfnamefont {M.}~\bibnamefont {Taylor}},
  \bibinfo {author} {\bibfnamefont {B.}~\bibnamefont {Weight}}, \bibinfo
  {author} {\bibfnamefont {E.}~\bibnamefont {Koessler}}, \bibinfo {author}
  {\bibfnamefont {X.}~\bibnamefont {Li}}, \ and\ \bibinfo {author}
  {\bibfnamefont {P.}~\bibnamefont {Huo}},\ }\href {\doibase
  10.26434/chemrxiv-2022-g9lr7} {\  (\bibinfo {year} {2022}),\
  10.26434/chemrxiv-2022-g9lr7}\BibitemShut {NoStop}%
\bibitem [{\citenamefont {Sidler}\ \emph {et~al.}(2022)\citenamefont {Sidler},
  \citenamefont {Ruggenthaler}, \citenamefont {Sch\"{a}fer}, \citenamefont
  {Ronca},\ and\ \citenamefont {Rubio}}]{Sidler2022}%
  \BibitemOpen
  \bibfield  {author} {\bibinfo {author} {\bibfnamefont {D.}~\bibnamefont
  {Sidler}}, \bibinfo {author} {\bibfnamefont {M.}~\bibnamefont
  {Ruggenthaler}}, \bibinfo {author} {\bibfnamefont {C.}~\bibnamefont
  {Sch\"{a}fer}}, \bibinfo {author} {\bibfnamefont {E.}~\bibnamefont {Ronca}},
  \ and\ \bibinfo {author} {\bibfnamefont {A.}~\bibnamefont {Rubio}},\ }\href
  {\doibase 10.1063/5.0094956} {\bibfield  {journal} {\bibinfo  {journal} {The
  Journal of Chemical Physics}\ }\textbf {\bibinfo {volume} {156}},\ \bibinfo
  {pages} {230901} (\bibinfo {year} {2022})}\BibitemShut {NoStop}%
\bibitem [{\citenamefont {Sukharev}\ and\ \citenamefont
  {Nitzan}(2017)}]{Sukharev2017}%
  \BibitemOpen
  \bibfield  {author} {\bibinfo {author} {\bibfnamefont {M.}~\bibnamefont
  {Sukharev}}\ and\ \bibinfo {author} {\bibfnamefont {A.}~\bibnamefont
  {Nitzan}},\ }\href {\doibase 10.1088/1361-648x/aa85ef} {\bibfield  {journal}
  {\bibinfo  {journal} {Journal of Physics: Condensed Matter}\ }\textbf
  {\bibinfo {volume} {29}},\ \bibinfo {pages} {443003} (\bibinfo {year}
  {2017})}\BibitemShut {NoStop}%
\bibitem [{\citenamefont {Scalora}\ \emph {et~al.}(2010)\citenamefont
  {Scalora}, \citenamefont {Vincenti}, \citenamefont {de~Ceglia}, \citenamefont
  {Roppo}, \citenamefont {Centini}, \citenamefont {Akozbek},\ and\
  \citenamefont {Bloemer}}]{Scalora2010}%
  \BibitemOpen
  \bibfield  {author} {\bibinfo {author} {\bibfnamefont {M.}~\bibnamefont
  {Scalora}}, \bibinfo {author} {\bibfnamefont {M.~A.}\ \bibnamefont
  {Vincenti}}, \bibinfo {author} {\bibfnamefont {D.}~\bibnamefont {de~Ceglia}},
  \bibinfo {author} {\bibfnamefont {V.}~\bibnamefont {Roppo}}, \bibinfo
  {author} {\bibfnamefont {M.}~\bibnamefont {Centini}}, \bibinfo {author}
  {\bibfnamefont {N.}~\bibnamefont {Akozbek}}, \ and\ \bibinfo {author}
  {\bibfnamefont {M.~J.}\ \bibnamefont {Bloemer}},\ }\href {\doibase
  10.1103/physreva.82.043828} {\bibfield  {journal} {\bibinfo  {journal}
  {Physical Review A}\ }\textbf {\bibinfo {volume} {82}} (\bibinfo {year}
  {2010}),\ 10.1103/physreva.82.043828}\BibitemShut {NoStop}%
\bibitem [{\citenamefont {Zeng}\ \emph {et~al.}(2009)\citenamefont {Zeng},
  \citenamefont {Hoyer}, \citenamefont {Liu}, \citenamefont {Koch},\ and\
  \citenamefont {Moloney}}]{Zeng2009}%
  \BibitemOpen
  \bibfield  {author} {\bibinfo {author} {\bibfnamefont {Y.}~\bibnamefont
  {Zeng}}, \bibinfo {author} {\bibfnamefont {W.}~\bibnamefont {Hoyer}},
  \bibinfo {author} {\bibfnamefont {J.}~\bibnamefont {Liu}}, \bibinfo {author}
  {\bibfnamefont {S.~W.}\ \bibnamefont {Koch}}, \ and\ \bibinfo {author}
  {\bibfnamefont {J.~V.}\ \bibnamefont {Moloney}},\ }\href {\doibase
  10.1103/PhysRevB.79.235109} {\bibfield  {journal} {\bibinfo  {journal} {Phys.
  Rev. B}\ }\textbf {\bibinfo {volume} {79}},\ \bibinfo {pages} {235109}
  (\bibinfo {year} {2009})}\BibitemShut {NoStop}%
\bibitem [{\citenamefont {Sukharev}\ \emph
  {et~al.}(2022{\natexlab{a}})\citenamefont {Sukharev}, \citenamefont
  {Drobnyh},\ and\ \citenamefont {Pachter}}]{Sukharev2022b}%
  \BibitemOpen
  \bibfield  {author} {\bibinfo {author} {\bibfnamefont {M.}~\bibnamefont
  {Sukharev}}, \bibinfo {author} {\bibfnamefont {E.}~\bibnamefont {Drobnyh}}, \
  and\ \bibinfo {author} {\bibfnamefont {R.}~\bibnamefont {Pachter}},\ }\href
  {\doibase 10.1063/5.0109872} {\bibfield  {journal} {\bibinfo  {journal} {The
  Journal of Chemical Physics}\ }\textbf {\bibinfo {volume} {157}},\ \bibinfo
  {pages} {134105} (\bibinfo {year} {2022}{\natexlab{a}})}\BibitemShut
  {NoStop}%
\bibitem [{\citenamefont {Cui}\ and\ \citenamefont {Nitzan}(2022)}]{Cui2022}%
  \BibitemOpen
  \bibfield  {author} {\bibinfo {author} {\bibfnamefont {B.}~\bibnamefont
  {Cui}}\ and\ \bibinfo {author} {\bibfnamefont {A.}~\bibnamefont {Nitzan}},\
  }\href {\doibase 10.1063/5.0101528} {\bibfield  {journal} {\bibinfo
  {journal} {The Journal of Chemical Physics}\ }\textbf {\bibinfo {volume}
  {157}},\ \bibinfo {pages} {114108} (\bibinfo {year} {2022})}\BibitemShut
  {NoStop}%
\bibitem [{\citenamefont {Luk}\ \emph {et~al.}(2017)\citenamefont {Luk},
  \citenamefont {Feist}, \citenamefont {Toppari},\ and\ \citenamefont
  {Groenhof}}]{Luk2017}%
  \BibitemOpen
  \bibfield  {author} {\bibinfo {author} {\bibfnamefont {H.~L.}\ \bibnamefont
  {Luk}}, \bibinfo {author} {\bibfnamefont {J.}~\bibnamefont {Feist}}, \bibinfo
  {author} {\bibfnamefont {J.~J.}\ \bibnamefont {Toppari}}, \ and\ \bibinfo
  {author} {\bibfnamefont {G.}~\bibnamefont {Groenhof}},\ }\href {\doibase
  10.1021/acs.jctc.7b00388} {\bibfield  {journal} {\bibinfo  {journal} {Journal
  of Chemical Theory and Computation}\ }\textbf {\bibinfo {volume} {13}},\
  \bibinfo {pages} {4324} (\bibinfo {year} {2017})}\BibitemShut {NoStop}%
\bibitem [{\citenamefont {Sukharev}\ \emph
  {et~al.}(2022{\natexlab{b}})\citenamefont {Sukharev}, \citenamefont
  {Subotnik},\ and\ \citenamefont {Nitzan}}]{sukharev2022}%
  \BibitemOpen
  \bibfield  {author} {\bibinfo {author} {\bibfnamefont {M.}~\bibnamefont
  {Sukharev}}, \bibinfo {author} {\bibfnamefont {J.}~\bibnamefont {Subotnik}},
  \ and\ \bibinfo {author} {\bibfnamefont {A.}~\bibnamefont {Nitzan}},\ }\href
  {\doibase 10.48550/ARXIV.2210.10943} {\enquote {\bibinfo {title}
  {Dissociation slowdown by collective optical response under strong coupling
  conditions},}\ } (\bibinfo {year} {2022}{\natexlab{b}})\BibitemShut {NoStop}%
\bibitem [{\citenamefont {Li}\ \emph {et~al.}(2021)\citenamefont {Li},
  \citenamefont {Nitzan},\ and\ \citenamefont {Subotnik}}]{Li2021}%
  \BibitemOpen
  \bibfield  {author} {\bibinfo {author} {\bibfnamefont {T.~E.}\ \bibnamefont
  {Li}}, \bibinfo {author} {\bibfnamefont {A.}~\bibnamefont {Nitzan}}, \ and\
  \bibinfo {author} {\bibfnamefont {J.~E.}\ \bibnamefont {Subotnik}},\ }\href
  {\doibase 10.1063/5.0037623} {\bibfield  {journal} {\bibinfo  {journal} {The
  Journal of Chemical Physics}\ }\textbf {\bibinfo {volume} {154}},\ \bibinfo
  {pages} {094124} (\bibinfo {year} {2021})}\BibitemShut {NoStop}%
\bibitem [{\citenamefont {Tavis}\ and\ \citenamefont
  {Cummings}(1968)}]{Tavis1968}%
  \BibitemOpen
  \bibfield  {author} {\bibinfo {author} {\bibfnamefont {M.}~\bibnamefont
  {Tavis}}\ and\ \bibinfo {author} {\bibfnamefont {F.~W.}\ \bibnamefont
  {Cummings}},\ }\href {\doibase 10.1103/PhysRev.170.379} {\bibfield  {journal}
  {\bibinfo  {journal} {Phys. Rev.}\ }\textbf {\bibinfo {volume} {170}},\
  \bibinfo {pages} {379} (\bibinfo {year} {1968})}\BibitemShut {NoStop}%
\bibitem [{\citenamefont {Tavis}\ and\ \citenamefont
  {Cummings}(1969)}]{Tavis1969}%
  \BibitemOpen
  \bibfield  {author} {\bibinfo {author} {\bibfnamefont {M.}~\bibnamefont
  {Tavis}}\ and\ \bibinfo {author} {\bibfnamefont {F.~W.}\ \bibnamefont
  {Cummings}},\ }\href {\doibase 10.1103/PhysRev.188.692} {\bibfield  {journal}
  {\bibinfo  {journal} {Phys. Rev.}\ }\textbf {\bibinfo {volume} {188}},\
  \bibinfo {pages} {692} (\bibinfo {year} {1969})}\BibitemShut {NoStop}%
\bibitem [{\citenamefont {Holstein}(1959{\natexlab{a}})}]{Holstein1959}%
  \BibitemOpen
  \bibfield  {author} {\bibinfo {author} {\bibfnamefont {T.}~\bibnamefont
  {Holstein}},\ }\href {\doibase https://doi.org/10.1016/0003-4916(59)90002-8}
  {\bibfield  {journal} {\bibinfo  {journal} {Ann. Phys.}\ }\textbf {\bibinfo
  {volume} {8}},\ \bibinfo {pages} {325} (\bibinfo {year}
  {1959}{\natexlab{a}})}\BibitemShut {NoStop}%
\bibitem [{\citenamefont {Holstein}(1959{\natexlab{b}})}]{Holstein1959b}%
  \BibitemOpen
  \bibfield  {author} {\bibinfo {author} {\bibfnamefont {T.}~\bibnamefont
  {Holstein}},\ }\href {\doibase https://doi.org/10.1016/0003-4916(59)90003-X}
  {\bibfield  {journal} {\bibinfo  {journal} {Ann. Phys.}\ }\textbf {\bibinfo
  {volume} {8}},\ \bibinfo {pages} {343} (\bibinfo {year}
  {1959}{\natexlab{b}})}\BibitemShut {NoStop}%
\bibitem [{\citenamefont {Spano}(2015)}]{Spano2015}%
  \BibitemOpen
  \bibfield  {author} {\bibinfo {author} {\bibfnamefont {F.~C.}\ \bibnamefont
  {Spano}},\ }\href {\doibase 10.1063/1.4919348} {\bibfield  {journal}
  {\bibinfo  {journal} {The Journal of Chemical Physics}\ }\textbf {\bibinfo
  {volume} {142}},\ \bibinfo {pages} {184707} (\bibinfo {year}
  {2015})}\BibitemShut {NoStop}%
\bibitem [{\citenamefont {Herrera}\ and\ \citenamefont
  {Spano}(2016)}]{Herrera2016}%
  \BibitemOpen
  \bibfield  {author} {\bibinfo {author} {\bibfnamefont {F.}~\bibnamefont
  {Herrera}}\ and\ \bibinfo {author} {\bibfnamefont {F.~C.}\ \bibnamefont
  {Spano}},\ }\href {\doibase 10.1103/PhysRevLett.116.238301} {\bibfield
  {journal} {\bibinfo  {journal} {Phys. Rev. Lett.}\ }\textbf {\bibinfo
  {volume} {116}},\ \bibinfo {pages} {238301} (\bibinfo {year}
  {2016})}\BibitemShut {NoStop}%
\bibitem [{\citenamefont {Galego}\ \emph {et~al.}(2015)\citenamefont {Galego},
  \citenamefont {Garcia-Vidal},\ and\ \citenamefont {Feist}}]{Galego2015}%
  \BibitemOpen
  \bibfield  {author} {\bibinfo {author} {\bibfnamefont {J.}~\bibnamefont
  {Galego}}, \bibinfo {author} {\bibfnamefont {F.~J.}\ \bibnamefont
  {Garcia-Vidal}}, \ and\ \bibinfo {author} {\bibfnamefont {J.}~\bibnamefont
  {Feist}},\ }\href {\doibase 10.1103/PhysRevX.5.041022} {\bibfield  {journal}
  {\bibinfo  {journal} {Phys. Rev. X}\ }\textbf {\bibinfo {volume} {5}},\
  \bibinfo {pages} {041022} (\bibinfo {year} {2015})}\BibitemShut {NoStop}%
\bibitem [{\citenamefont {Galego}\ \emph {et~al.}(2016)\citenamefont {Galego},
  \citenamefont {Garcia-Vidal},\ and\ \citenamefont {Feist}}]{Galego2016b}%
  \BibitemOpen
  \bibfield  {author} {\bibinfo {author} {\bibfnamefont {J.}~\bibnamefont
  {Galego}}, \bibinfo {author} {\bibfnamefont {F.~J.}\ \bibnamefont
  {Garcia-Vidal}}, \ and\ \bibinfo {author} {\bibfnamefont {J.}~\bibnamefont
  {Feist}},\ }\href {\doibase 10.1038/ncomms13841} {\bibfield  {journal}
  {\bibinfo  {journal} {Nature Communications}\ }\textbf {\bibinfo {volume}
  {7}} (\bibinfo {year} {2016}),\ 10.1038/ncomms13841}\BibitemShut {NoStop}%
\bibitem [{\citenamefont {Wu}\ \emph {et~al.}(2016)\citenamefont {Wu},
  \citenamefont {Feist},\ and\ \citenamefont {Garcia-Vidal}}]{Wu2016}%
  \BibitemOpen
  \bibfield  {author} {\bibinfo {author} {\bibfnamefont {N.}~\bibnamefont
  {Wu}}, \bibinfo {author} {\bibfnamefont {J.}~\bibnamefont {Feist}}, \ and\
  \bibinfo {author} {\bibfnamefont {F.~J.}\ \bibnamefont {Garcia-Vidal}},\
  }\href {\doibase 10.1103/PhysRevB.94.195409} {\bibfield  {journal} {\bibinfo
  {journal} {Phys. Rev. B}\ }\textbf {\bibinfo {volume} {94}},\ \bibinfo
  {pages} {195409} (\bibinfo {year} {2016})}\BibitemShut {NoStop}%
\bibitem [{\citenamefont {Zeb}\ \emph {et~al.}(2017)\citenamefont {Zeb},
  \citenamefont {Kirton},\ and\ \citenamefont {Keeling}}]{Zeb2017}%
  \BibitemOpen
  \bibfield  {author} {\bibinfo {author} {\bibfnamefont {M.~A.}\ \bibnamefont
  {Zeb}}, \bibinfo {author} {\bibfnamefont {P.~G.}\ \bibnamefont {Kirton}}, \
  and\ \bibinfo {author} {\bibfnamefont {J.}~\bibnamefont {Keeling}},\ }\href
  {\doibase 10.1021/acsphotonics.7b00916} {\bibfield  {journal} {\bibinfo
  {journal} {{ACS} Photonics}\ }\textbf {\bibinfo {volume} {5}},\ \bibinfo
  {pages} {249} (\bibinfo {year} {2017})}\BibitemShut {NoStop}%
\bibitem [{Note1()}]{Note1}%
  \BibitemOpen
  \bibinfo {note} {Note that in the 1-exciton approximation, we can replace the
  Hamiltonian \protect \textup {\hbox {\mathsurround \z@ \protect \normalfont
  (\ignorespaces \ref {eq:HHTC}\unskip \@@italiccorr )}} by its rotating state
  version, \begin {equation} \protect \hat {H}_{TC-RWA}=\hbar \omega _c\protect
  \hat {a}^\dagger \protect \hat {a}+\hbar \DOTSB \sum@ \slimits@ _{j=1}^N\left
  [\omega _{xg}\protect \hat {\sigma }^+_j\protect \hat {\sigma }^-_j+\protect
  \frac {g}{2}(\protect \hat {a}^\dagger \protect \hat {\sigma }^-_j+\protect
  \hat {a}\protect \hat {\sigma }^+_j)\right ], \label {eq:HHTCrwa} \end
  {equation}}\BibitemShut {NoStop}%
\bibitem [{Note2()}]{Note2}%
  \BibitemOpen
  \bibinfo {note} {In the mean-field representation used in models TC2 and CN2,
  the molecular bright state is taken by letting $c_{je}=N^{-1/2},
  c_{jg}=(1-N^{-1})^{1/2}$ in Eq. \protect \textup {\hbox {\mathsurround \z@
  \protect \normalfont (\ignorespaces \ref {eq:waveftrunc}\unskip \@@italiccorr
  )}} or $c_{jea}=N^{-1/2},c_{jga}=(1-N^{-1})^{1/2}$ in Eq. \protect \textup
  {\hbox {\mathsurround \z@ \protect \normalfont (\ignorespaces \ref
  {eq:wavefsemi}\unskip \@@italiccorr )}}.}\BibitemShut {Stop}%
\bibitem [{Note3()}]{Note3}%
  \BibitemOpen
  \bibinfo {note} {Taking $\eta _a\protect \neq 0$ corresponds to relaxation
  processes that do not lead to this product}\BibitemShut {NoStop}%
\bibitem [{\citenamefont {T\"{o}rm\"{a}}\ and\ \citenamefont
  {Barnes}(2014)}]{Trm2014}%
  \BibitemOpen
  \bibfield  {author} {\bibinfo {author} {\bibfnamefont {P.}~\bibnamefont
  {T\"{o}rm\"{a}}}\ and\ \bibinfo {author} {\bibfnamefont {W.~L.}\ \bibnamefont
  {Barnes}},\ }\href {\doibase 10.1088/0034-4885/78/1/013901} {\bibfield
  {journal} {\bibinfo  {journal} {Reports on Progress in Physics}\ }\textbf
  {\bibinfo {volume} {78}},\ \bibinfo {pages} {013901} (\bibinfo {year}
  {2014})}\BibitemShut {NoStop}%
\bibitem [{\citenamefont {Campaioli}\ \emph {et~al.}(2019)\citenamefont
  {Campaioli}, \citenamefont {Pollock},\ and\ \citenamefont
  {Vinjanampathy}}]{Campaioli2018}%
  \BibitemOpen
  \bibfield  {author} {\bibinfo {author} {\bibfnamefont {F.}~\bibnamefont
  {Campaioli}}, \bibinfo {author} {\bibfnamefont {F.}~\bibnamefont {Pollock}},
  \ and\ \bibinfo {author} {\bibfnamefont {S.}~\bibnamefont {Vinjanampathy}},\
  }\enquote {\bibinfo {title} {Quantum batteries},}\ in\ \href {\doibase
  10.1007/978-3-319-99046-0_8} {\emph {\bibinfo {booktitle} {Thermodynamics in
  the Quantum Regime}}},\ \bibinfo {series} {Fundamental Theories of Physics},
  Vol.\ \bibinfo {volume} {195},\ \bibinfo {editor} {edited by\ \bibinfo
  {editor} {\bibfnamefont {F.}~\bibnamefont {Binder}}, \bibinfo {editor}
  {\bibfnamefont {L.}~\bibnamefont {Correa}}, \bibinfo {editor} {\bibfnamefont
  {C.}~\bibnamefont {Gogolin}}, \bibinfo {editor} {\bibfnamefont
  {J.}~\bibnamefont {Anders}}, \ and\ \bibinfo {editor} {\bibfnamefont
  {G.}~\bibnamefont {Adesso}}}\ (\bibinfo  {publisher} {Springer},\ \bibinfo
  {year} {2019})\ pp.\ \bibinfo {pages} {207--225}\BibitemShut {NoStop}%
\bibitem [{\citenamefont {Bhattacharjee}\ and\ \citenamefont
  {Dutta}(2021)}]{Bhattacharjee2021}%
  \BibitemOpen
  \bibfield  {author} {\bibinfo {author} {\bibfnamefont {S.}~\bibnamefont
  {Bhattacharjee}}\ and\ \bibinfo {author} {\bibfnamefont {A.}~\bibnamefont
  {Dutta}},\ }\href {\doibase 10.1140/epjb/s10051-021-00235-3} {\bibfield
  {journal} {\bibinfo  {journal} {The European Physical Journal B}\ }\textbf
  {\bibinfo {volume} {94}} (\bibinfo {year} {2021}),\
  10.1140/epjb/s10051-021-00235-3}\BibitemShut {NoStop}%
\end{thebibliography}%

\end{document}

\title{Supplementary Information for "Comparing semiclassical mean-field and 1-exciton approximations in evaluating optical response under strong light-matter coupling conditions"}

\author{Bingyu Cui}
\affiliation{Department of Chemistry, University of Pennsylvania, Philadelphia, Pennsylvania 19104,
USA}
\affiliation{School of Chemistry, Tel Aviv University, Tel Aviv 69978, Israel}

\author{Maxim Sukharev}
\affiliation{Department of Physics, Arizona State University, Tempe, Arizona 85287, USA}
\affiliation{College of Integrative Sciences and Arts, Arizona State University, Mesa, Arizona 85201, USA}

\author{Abraham Nitzan}
\email{anitzan@sas.upenn.edu}
\affiliation{Department of Chemistry, University of Pennsylvania, Philadelphia, Pennsylvania 19104,
USA}
\affiliation{School of Chemistry, Tel Aviv University, Tel Aviv 69978, Israel}

\date{\today}

\pacs{}
\maketitle

\section{the self-consistent field approximation}
\setcounter{equation}{0}
\label{app.1}
In this section, we discuss the Hartree level approximation for the molecular wavefunciton in the TC2 and CN2 models. It suffices to consider a pair of molecules without internal (e.g. vibrational) states. The generalization to many molecules and to models with internal levels is straightforward. The full Hamiltonian of two molecules interacting with the cavity mode represented by a classical oscillator is
\begin{equation}
    H=\frac{p^2+\omega_c^2x^2}{2}+\hbar\omega_{xg}\hat{\sigma}^+_1\hat{\sigma}^-_1+\hbar\omega_{xg}\hat{\sigma}^+_2\hat{\sigma}^-_2+\frac{\hbar g}{2}\sqrt{\frac{2\omega_c}{\hbar}}x(\hat{\sigma}_1^-+\hat{\sigma}_1^++\hat{\sigma}_2^-+\hat{\sigma}_2^+),
    \label{eq:semiTC}
\end{equation}
where $\hat{\sigma}_i^+=|e_i\rangle\langle g_i|,\hat{\sigma}_i^-=|g_i\rangle\langle e_i|,i=1,2$. The molecular wavefunction $\Psi(t)$ is a vector in the Hilbert space $\mathcal{H}_1\otimes\mathcal{H}_2$, satisfying the Schr\"{o}dinger equation $i\hbar d\Psi/dt=\hat{H}\Psi$. It can be expressed as 
\begin{equation}
\Psi(t)=\psi_1(t)\otimes\psi_2(t)=c_{11}(t)|g_1\rangle\otimes|g_2\rangle+c_{12}(t)|g_1\rangle\otimes|e_2\rangle+c_{21}(t)|e_1\rangle\otimes|g_2\rangle+c_{22}(t)|e_1\rangle\otimes|e_2\rangle,
\end{equation}
where $c_{11}(t),c_{12}(t),c_{21}(t),c_{22}(t)$ are time-dependent complex coefficients. 
%Note that the molecular wavefunction $\Psi(t)$ is normalized,
%\begin{equation}
%    \frac{d\langle\Psi|\Psi\rangle}{dt}=\langle\dot{\Psi}|\Psi\rangle+\langle\Psi|\dot{\Psi}\rangle=\frac{i}{\hbar}\langle\Psi|H|\Psi\rangle-\frac{i}{\hbar}\langle\Psi|H|\Psi\rangle=0.
%\end{equation}

Meanwhile, the Hamiltonian for the classical harmonic cavity mode is obtained from
\begin{align}
   \tilde{H}&=\langle\Psi(t)|\hat{H}|\Psi(t)\rangle.
\end{align}
This leads to the classical equations of the motion for the classical mode in the form
\begin{align}
   \begin{cases}
   \dot{p}=-\frac{\partial\tilde{H}}{\partial x}=-\omega_c^2x-\frac{\hbar g}{2}\sqrt{\frac{2\omega_c}{\hbar}}(c_{11}^*c_{21}+c_{11}^*c_{12}+c_{21}^*c_{22}+c_{12}^*c_{22}+c_{21}^*c_{11}+c_{12}^*c_{11}+c_{22}^*c_{21}+c_{22}^*c_{12})\\
   \dot{x}=\frac{\partial\tilde{H}}{\partial p}=p.
   \end{cases}    
   \label{eq:eomclassical}
\end{align}

Note that this Ehrenfest dynamics conserves the total energy:
\begin{align}
    \frac{d\tilde{H}}{dt}&=\langle\frac{d\Psi}{dt}|H|\Psi\rangle+\langle\Psi|H|\frac{d\Psi}{dt}\rangle+\langle\Psi|\frac{dH}{dt}|\Psi\rangle\notag\\
    &=p\dot{p}+\omega_c^2x\dot{x}+\frac{\hbar g}{2}\sqrt{\frac{2\omega_c}{\hbar}}\dot{x}\langle\Psi|\hat{\sigma}_1^++\hat{\sigma}_1^-+\hat{\sigma}_2^++\hat{\sigma}_2^-|\Psi\rangle\notag\\
    &=0,
    \label{eq:energyconserve}
\end{align}
where the sum of the first two terms on the 1st line vanishes.
 
If the total wavefunction can be written as a product of normalized wavefunctions, $(\psi_i,i=1,2)$, each describing the state of a single atom, 
\begin{equation}
    \Psi(t)=\psi_1(t)\psi_2(t),
\end{equation}
substituting it back to Eq. \eqref{eq:semiTC} and denoting 
\begin{equation}
    \hat{V}_1=\frac{\hbar g}{2}\sqrt{\frac{2\omega_c}{\hbar}}x(\hat{\sigma}_1^-+\hat{\sigma}_1^+);\quad  \hat{V}_2=\frac{\hbar g}{2}\sqrt{\frac{2\omega_c}{\hbar}}x(\hat{\sigma}_2^-+\hat{\sigma}_2^+),
\end{equation}
we obtain
\begin{equation}
    i\hbar\left(\psi_2\frac{\partial \psi_1}{\partial t}+\psi_1\frac{\partial \psi_2}{\partial t}\right)=\left(\frac{p^2+\omega_c^2x^2}{2}\right)\psi_1\psi_2+\psi_2\hat{H}_1\psi_1+\psi_1\hat{H}_2\psi_2+(\hat{V}_1+\hat{V}_2)\psi_1\psi_2.
    \label{eq:Subs}
\end{equation}
where $\hat{H}_i\equiv\hbar\omega_{xg}\hat{\sigma}^+_i\hat{\sigma}^-_i,i=1,2$. Multiplying Eq. \eqref{eq:Subs} by $\psi_1^*$ and integrating over the coordinates of subsystem associated with molecule 2, or by $\psi_2^*$ and integrating over the coordinates of molecule 1, we have,
\begin{subequations}
\begin{align}
    \label{eq:necea}
    i\hbar\frac{\partial \psi_1}{\partial t}&=\epsilon_2(t)\psi_1+\left(\frac{p^2+\omega_c^2x^2}{2}\right)\psi_1+\hat{H}_1\psi_1+\langle \psi_2|\hat{V}|\psi_2\rangle\psi_1,\\
    i\hbar\frac{\partial \psi_2}{\partial t}&=\epsilon_1(t)\psi_2+\left(\frac{p^2+\omega_c^2x^2}{2}\right)\psi_2+\hat{H}_2\psi_2+\langle \psi_1|\hat{V}|\psi_1\rangle\psi_2,\\
    \epsilon_1(t)&=-i\hbar\langle\psi_1|\frac{\partial \psi_1}{\partial t}\rangle+\langle\psi_1|\hat{H}_1|\psi_1\rangle,\\
    \epsilon_2(t)&=-i\hbar\langle\psi_2|\frac{\partial \psi_2}{\partial t}\rangle+\langle\psi_2|\hat{H}_2|\psi_2\rangle.   
\end{align}
\label{eq:nece}
\end{subequations}
The equations of motion for the classical oscillator remain the same as Eq. \eqref{eq:eomclassical}.

To continue, assume that the two wavefunctions $\psi_1$ and $\psi_2$ satisfy
\begin{equation}
    i\hbar\dot{\psi}_1=\left(\frac{p^2+\omega_c^2x^2}{4}\right)\psi_1+\hat{H}_1\psi_1+\hat{V}_1\psi_1;  \quad i\hbar\dot{\psi}_2=\left(\frac{p^2+\omega_c^2x^2}{4}\right)\psi_2+\hat{H}_2\psi_2+\hat{V}_2\psi_2,
    \label{eq:suff}
\end{equation}
then,
\begin{align}
    i\hbar(\dot{\psi}_1\psi_2+\psi_1\dot{\psi}_2)=\left(\frac{p^2+\omega_c^2x^2}{4}\right)(\psi_1\psi_2+\psi_2\psi_1)+\hat{H}_1\psi_1\psi_2+\hat{V}_1\psi_1\psi_2+\hat{H}_2\psi_2\psi_1+\hat{V}_2\psi_2\psi_1
\end{align}
Thus, we can define $\Psi=\psi_1\psi_2=\psi_2\psi_1$, which solves the Schr\"{o}dinger equation associated with the Hamiltonian \eqref{eq:semiTC},
\begin{equation}
    i\hbar\frac{\partial \Psi}{\partial t}=\hat{H}\Psi.
\end{equation}

%Compare Eq. \eqref{eq:nece} and Eq. \eqref{eq:suff}, it looks that these are different equations for $\psi_1$ and $\psi_2$, obtained from the necessary condition and sufficient condition, respectively. However the uniqueness and existence of the linear differential equations requires they should be equal to each other. In other words, 
If we calculate $\psi_1,\psi_2$ from Eq. \eqref{eq:suff}, as well as the equation of motion for the classical oscillator, then they would also satisfy Eq. \eqref{eq:nece}. Indeed, assume Eq. \eqref{eq:suff}, then for Eq. (\ref{eq:necea}), we have
\begin{align}
    RHS&=\epsilon_2(t)\psi_1+\left(\frac{p^2+\omega_c^2x^2}{2}\right)\psi_1+\hat{H}_1\psi_1+\langle \psi_2|\hat{V}|\psi_2\rangle\psi_1\notag\\
    &=\left(\frac{p^2+\omega_c^2x^2}{4}\right)\psi_1\psi_1+\hat{H}_1\psi_1+\hat{V}_1\psi_1\notag\\
    &+\left[-i\hbar\langle\psi_2|\frac{\partial \psi_2}{\partial t}\rangle+\langle\psi_2|\hat{H}_2|\psi_2\rangle+\left(\frac{p^2+\omega_c^2x^2}{4}\right)+\langle\psi_2|\hat{V}_1|\psi_2\rangle\right]\psi_1\notag\\
    &=i\hbar\frac{\partial \psi_1}{\partial t}\notag\\
    &=LHS.
\end{align}
Terms in the mid-bracket vanish because of Eq. \eqref{eq:suff}. A similar identity holds for $\psi_2$. Therefore, $\Psi=\psi_1\psi_2=\psi_2\psi_1$ where $\psi_1,\psi_2$ solve Eq. \eqref{eq:suff}, is the solution of the Hamiltonian of two atoms interacting with a classical oscillator, subject to proper initial conditions.

\section{include damping terms}
\setcounter{equation}{14}
\label{app.2}
In this section, we sketch the way to add damping rates to states of interest in the CN1 (and TC1) model. In the truncated basis $|j,k\rangle\equiv|X_j\rangle|V_k\rangle$, the Hamiltonian (12) in the main text might be represented by an $(N+1)^2\times(N+1)^2$ matrix as the tensor product of $(N+1)$-dimensional (square) matrix representing electronic subspace (1-exciton TC Hamiltonian) spanned by the basis $|X_j\rangle,j=0,...,N$ and the $N+1$-dimensional (square) matrix of internal subspace spanned by the states $|V_k\rangle,k=0,...,N$, 
\begin{align}
    \mathbf{H}&=\left(\begin{matrix}
    \mathbf{M}^{(00)} &\mathbf{M}^{(01)} &... &\mathbf{M}^{(0N)}\\
    \mathbf{M}^{(10)} &\mathbf{M}^{(11)} &... &\mathbf{M}^{(1N)}\\
    &...\\
    \mathbf{M}^{(N0)} &\mathbf{M}^{(N1)} &... &\mathbf{M}^{(NN)}\\
\end{matrix}\right),
 \label{eq:matrixETC}
\end{align}
where each (square) block $\mathbf{M}$ is of order $N+1$. For diagonal block matrices, we have
\begin{align}
    \hbar^{-1}\mathbf{M}^{(jj)}&=\left(\begin{matrix}
    \omega_c+\Delta \omega(1-\delta_{j0}) &\frac{g}{2} &\frac{g}{2} &... &\frac{g}{2}\\
    \frac{g}{2} &\omega_{xg}+\Delta \omega(1-\delta_{j0}) &0 &... &0\\
    \frac{g}{2} &0 &\omega_{xg}+\Delta \omega(1-\delta_{j0}) &... &0\\
    &...\\
    \frac{g}{2} &0 &... &0 &\omega_{xg}+\Delta \omega(1-\delta_{j0})
\end{matrix}\right)\notag\\
j&=0,...,N.
\end{align}
For off diagonal elements, all matrices like $\mathbf{M}^{jj'}$ with $j\neq j'$ are 0 if both $j,j'$ are non zero, while the matrices $\mathbf{M}^{(0j)}$ and $\mathbf{M}^{(j0)}, j=1,...,N,$ have only one non-zero element  $\mathbf{M}^{(0j)}_{kl}=\mathbf{M}^{(j0)}_{kl}=\lambda\delta_{k,j+1}\delta_{l,j+1}$. The diagonal matrix of damping coefficients $\pmb{\eta}$ can be constructed according to states associated with damping fluxes. For example, if we want to open the damping channel for the cavity mode $|X_0\rangle|V_k\rangle,k=0,...,N$, then only entries $(1+k(N+1),1+k(N+1))$ in $\pmb{\eta}$ are non-zero and take value $\eta_c$. In other words, the matrix form of $\pmb{\eta}$ reads, in this case,
\begin{align}
    \pmb{\eta}&=\left(\begin{matrix}
    \pmb{\xi}^{(0)} &\mathbf{0} &... &\mathbf{0}\\
    \mathbf{0} &\pmb{\xi}^{(1)} &... &\mathbf{0}\\
    &...\\
    \mathbf{0} &\mathbf{0} &... &\pmb{\xi}^{(N)}\\
\end{matrix}\right),
 \label{eq:matrixETC}
\end{align}
where each block $\pmb{\xi}$ (and zero matrix $\mathbf{0}$) is of order $N+1$ and takes form
\begin{align}
    \pmb{\xi}^{(j)}&=\left(\begin{matrix}
    \eta_c &0 &0 &... &0\\
    0 &0 &0 &... &0\\
    0 &0 &0 &... &0\\
    &...\\
    0 &0 &... &0 &0
\end{matrix}\right), j=0,...,N.
\end{align}

\section{Additional numerical results}
\setcounter{equation}{18}
\label{app.3}

\begin{figure}[!htp]
\subfloat[][]{
\begin{minipage}[t]{0.5\textwidth}
\flushleft
\includegraphics[width=0.8\textwidth]{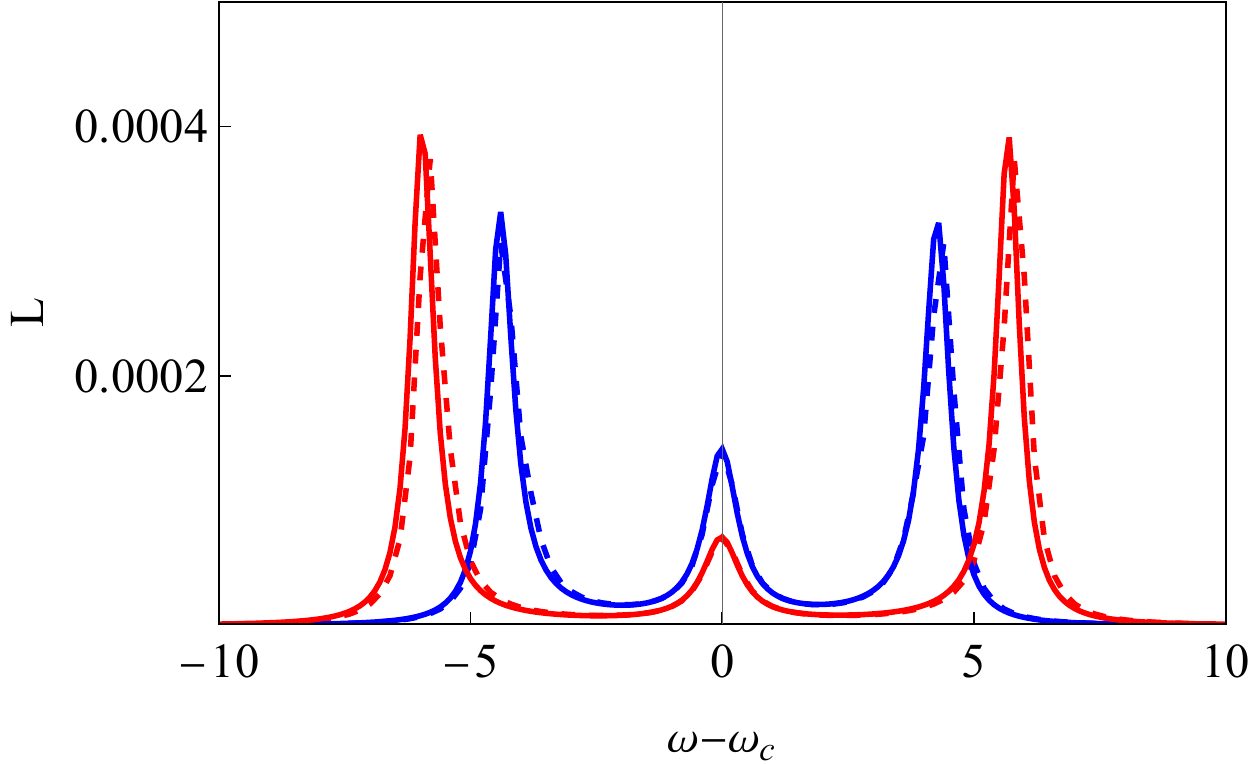}
\end{minipage}
}
\\
\subfloat[][]{
\begin{minipage}[t]{0.5\textwidth}
\flushleft
\includegraphics[width=0.8\textwidth]{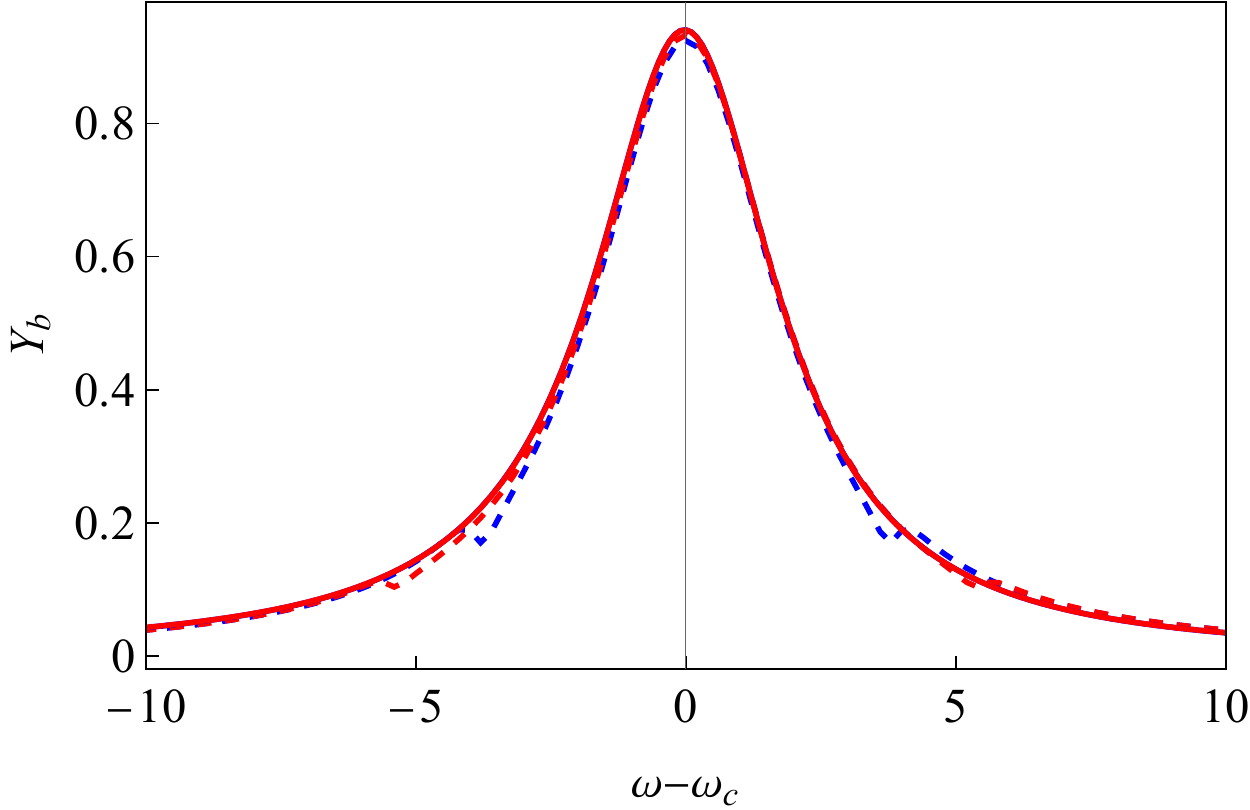}
\end{minipage}
}
\caption{(a) Absorption lineshapes (arbitrary units), plotted against the pumping frequency. The lineshapes calculated from Eqs. (21) and (24) in the main text are scaled to have the same height of the middle peak, (b) The yield $Y_b$, calculated from the CN1 (Eq. (22) in the main text, dashed lines),  and CN2 (Eq. (25) in the main text, solid lines) models, for molecular clusters of size $N=60$ (blue) or $N=120$ (red). Damping channels are opened for excited molecules in inner state $b$ and for the cavity mode/classical oscillator with damping coefficients $\eta_a=\eta_c=g$, respectively, while $\eta_c=0$. Other parameters are $g=0.1,\hbar\omega_c=E_{xg}=100g,W=0.01g, A=0.01, \Delta\omega=0$ and $\lambda=2g$.}
\label{fig:S1}
\end{figure}

\begin{figure}[!htp]
\subfloat[][]{
\begin{minipage}[t]{0.5\textwidth}
\flushleft
\includegraphics[width=0.8\textwidth]{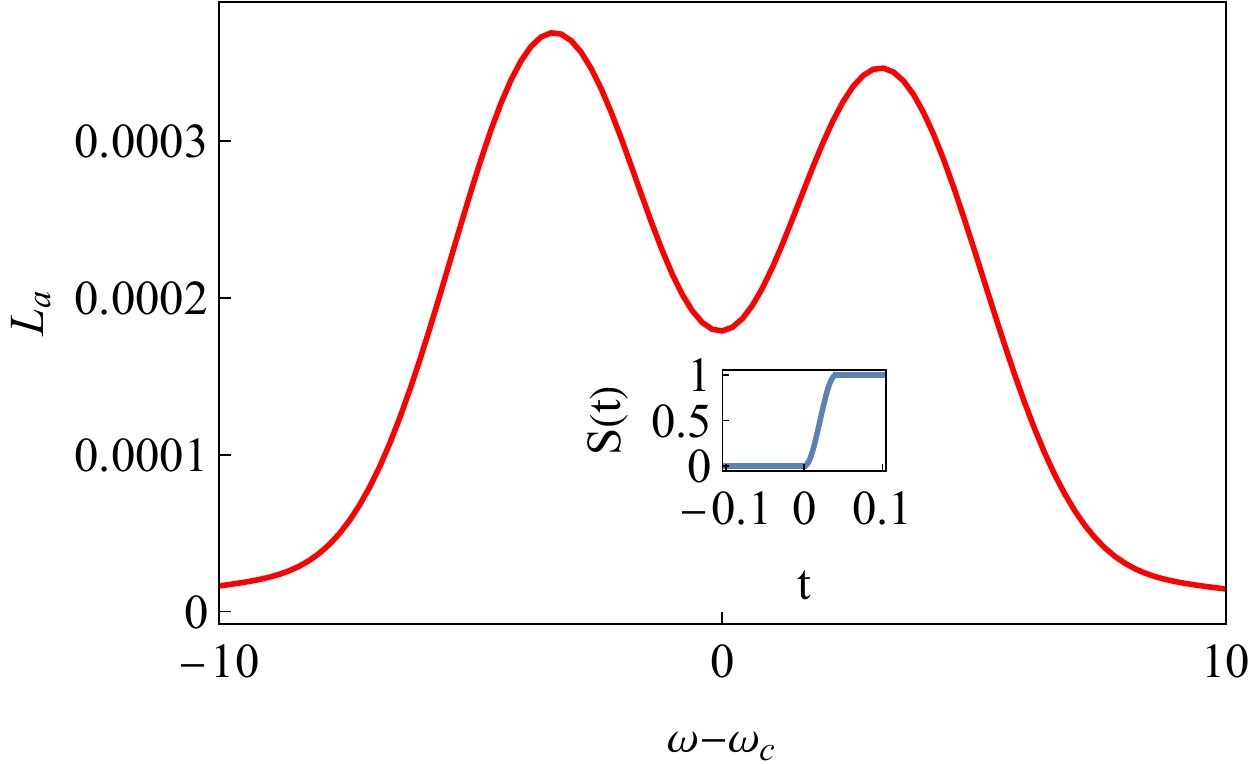}
\end{minipage}
}
\\
\subfloat[][]{
\begin{minipage}[t]{0.5\textwidth}
\flushleft
\includegraphics[width=0.8\textwidth]{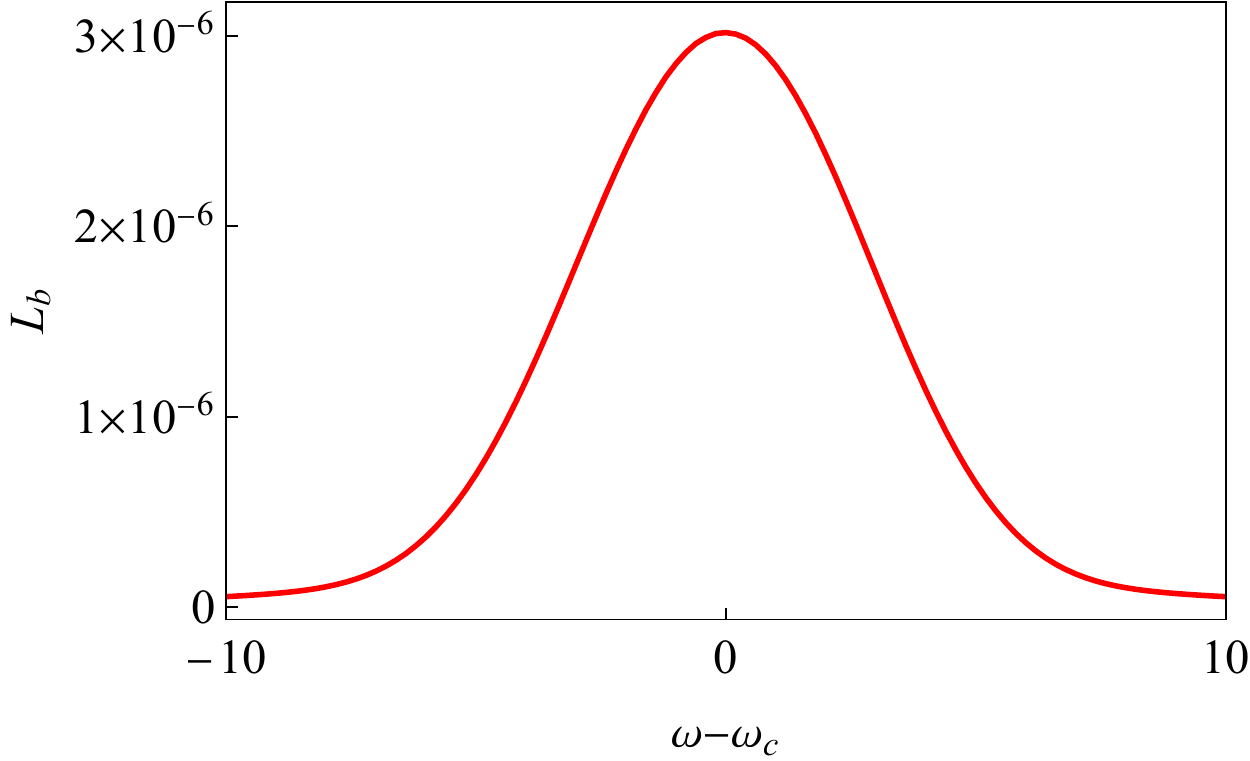}
\end{minipage}
}
\\
\subfloat[][]{
\begin{minipage}[t]{0.5\textwidth}
\flushleft
\includegraphics[width=0.8\textwidth]{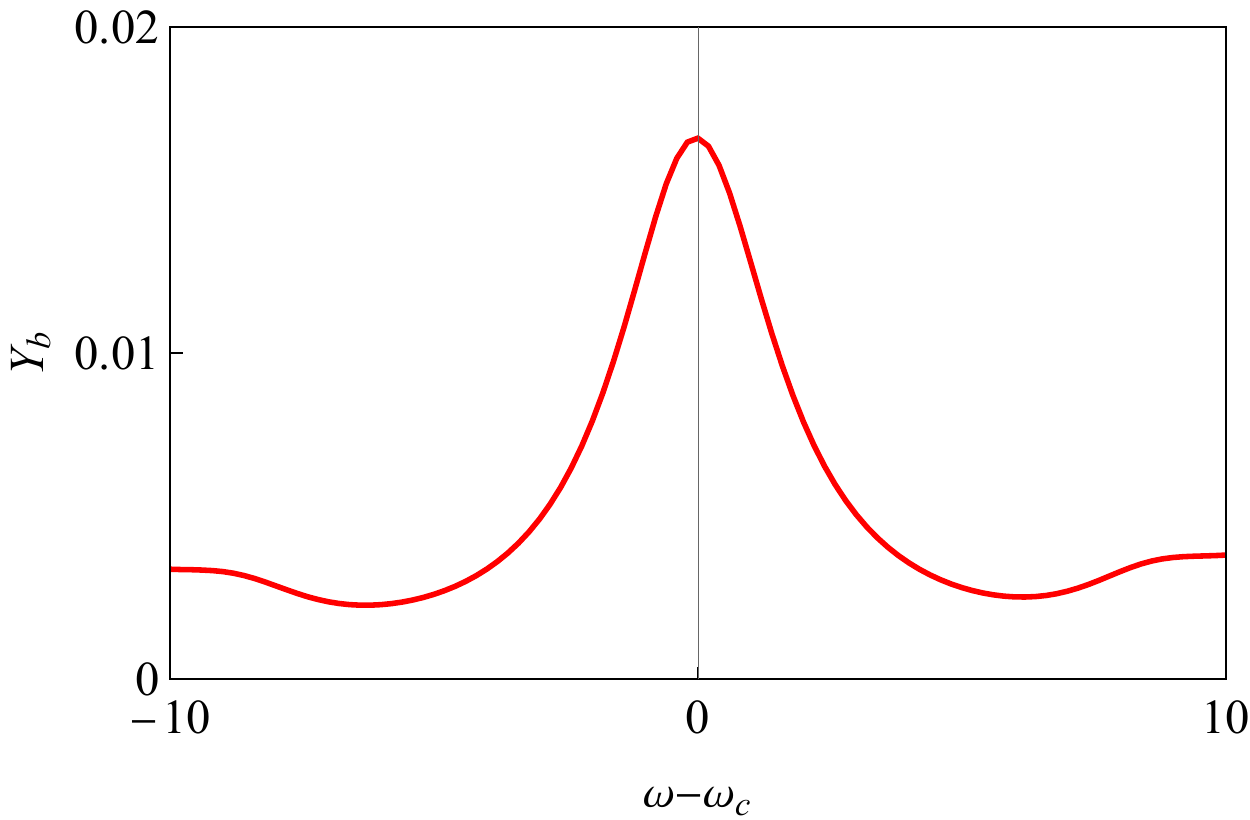}
\end{minipage}
}
\caption{Fluxes and yield obtained for the CN2 model when the system is under the switching field. Panel (a): flux through the cavity mode represented by the classical oscillator, the inset is the shape of the transient field, c.f., Eq. (26) in the main text; panel (b): flux through inner states $b$ in excited molecules; panel (c): yield of inner states $b$, c.f., Eq. (27d) in the main text. Before the switching, there are $N=60$ molecules in the $a$-ground state. The characteristic time of the pulse envelope $t_s$ is $\pi/(5g\sqrt{N})$. The damping coefficients are $\eta_a=\eta_b=g, \eta_c=0$. The snapshot is taken at $t=20t_s$. Other parameters are $g=0.1, \hbar\omega_c=E_{xg}=100g, A=0.01, \Delta\omega=0$ and $\lambda=0.2g$.}
\label{fig:S2}
\end{figure}

In this section, we present more figures as complements for those presented in Section III in the main text. 

In Fig. \ref{fig:S1}, we show results of lineshape and yield similar to Fig. 5 in the main text except that in addition to damping on inner state $b$ in excited molecules, a loss channel is placed on the inner state $a$ of excited molecules rather than on the cavity mode/classical oscillator.

Figure \ref{fig:S2} shows energy fluxes and yield of molecular response to the switching field at short time when the damping channels are opened for excited molecules in inner states $a$ and $b$.

\bibliography{reference}